\documentclass[fleqn,twoside]{article}%
\usepackage{amsmath}
\usepackage{amsfonts}
\usepackage{amssymb}
\usepackage{graphicx}
\usepackage{caption}
\usepackage{cite}

\def\be{\begin{equation}}
\def\ee{\end{equation}}
\def\bi{\bibitem}
\begin{document}
\title{Inflation and cosmological evolution with F(R,G) gravity theory.}
\author{Dalia Saha$^1$, Jyoti Prasad Saha$^2$ and Abhik Kumar Sanyal$^3$}
\maketitle
\noindent
\begin{center}
\noindent
$^{1,3}$ Dept. of Physics, Jangipur College, Murshidabad, West Bengal, India - 742213,\\
$^{1,2}$ Dept. of Physics, University of Kalyani, Nadia, West Bengal, India - 741235.\\
\end{center}
\footnotetext[1]{
\noindent Electronic address:\\
$^1$daliasahamandal1983@gmail.com\\
$^2$jyotiprasadsaha@gmail.com\\
$^3$sanyal\_ ak@yahoo.com\\}
\begin{abstract}
In the last decade Planck PR4 data together with ground-based experimental data such as, BK18, BAO and CMB lensing, tightened constraint of the tensor to scalar ratio, starting form $r < 0.14$ to $r < .032$, while the spectral index lies within the range $0.9631 \le n_s \le 0.9705$. Viability of modified gravity theories, proposed as alternatives to the dark-energy issue, should therefore be tested in the light of such new result. Here, we explore F(R, G) gravity theory in regard of the early universe and have shown that, it is not compatible with newly released constraints on $r$ and $n_s$ simultaneously. Further, it also fails to produce a feasible radiation dominated era. It therefore questions the justification of using the model for resolving the cosmic puzzle.
\end{abstract}
\section{Introduction:-}

The challenge to explain the issue of dark energy is now almost three decades old. In the first decade, cosmologists tinkered with different exotic dark energy models. However despite tremendous effort, no trace of phi (the non-interacting scalar field) has yet been found in the sky. Cosmologists now started believing that the puzzle may be resolved by modifying the left hand side of the Einstein's equation, namely the geometric part of the theory. The problem is, as for different dark energy models, almost all the modified theories of gravity can associate acceleration at the late-stage of cosmic evolution. It is therefore required to test such models from different astrophysical observations. In recent years, we have taken up the task to justify validity of these modified gravity theories, by studying these, in different cosmological regimes.\\

Out of a large number of different modified theories of gravity that have been put forward over decades to replace dark energy issue, $F(R,\mathcal{G})$ model \cite{I1,I2,I3,I4,I5,I6,I7,I8,I9,I10} is predominant and also prevalent, where, $\mathcal{G} = R^2 - 4 R_{\alpha\beta} R^{\alpha\beta} + R_{\alpha\beta\gamma\delta}R^{\alpha\beta\gamma\delta}$, being the Gauss-Bonnet term. The primary importance of such a model is that it does not require a scalar field to explain cosmic puzzle at the late stage of its evolution, provided non-linear term involving $\mathcal{G}$ is taken into account. The model has been scrutinized over years from different perspectives and supposed to have passed constraints, both in the ultraviolet and the infrared regimes, successfully \cite{I5,I11,I12,I13,I14,I15,I16,I17,I18,I20,I21,I22}. The feasibility of $F(R,\mathcal{G})$ gravity theory has also been demonstrated in a large number of articles \cite{I1,I2,I3,I4,I5,I6,I7,I8,I9,I10,I11,I12,I13,I14,I15,I79,I80,I81} Further, stability criteria is derived in view of the one-loop effective action of $F(\mathcal{G})$ gravity in de Sitter background \cite{I82}. Since the theory does not manifest any correction to Newton’s law in flat space on one hand, and also exhibits stability on the other, choosing $F(\mathcal{G})$ arbitrarily, so it is absolutely acquiescent with the solar-system constraints.  Also, it was demonstrated that the choice $F(\mathcal{G}) = a_0 \mathcal{G}^{n_1} + b_0 \mathcal{G}^{n_2}$ can unify inflation (if $n_1 > 1$) with the late-time accelerated expansion, (if $n_2 < {1\over 2}$) \cite{I83}. Considering all these investigations, it appears that $F(R, \mathcal{G})$ gravity may be a satisfactory alternative to the dark energy. Our aim is this manuscript is to study cosmic evolution with a generalized $F(R, \mathcal{G})$ theory of gravity. As the following action

\be\label{A1} A_1 = \int \left[\alpha R + \gamma \mathcal{G}^2\right]\sqrt{-g} d^4 x,\ee
was examined in the context of the evolution of early universe, a plethora of pathologies appear. Some of those were manoeuvered  including of a bare cosmological constant term \cite{I23}. Further, from the point of view of quantum cosmology, it has recently been noticed that one can not accommodate in the Einstein-Hilbert action, (or in any linear minimally or non-minimally coupled action) higher order terms of different orders \cite{I24}. For example, in the following action with field dependent coupling parameters $\alpha(\phi)$, $\beta(\phi)$ and $\gamma(\phi)$, and a cosmological constant $\Lambda$,

\be \label{A2} A_2 = \int \left[\alpha(\phi)(R - 2\Lambda) +\beta(\phi)R^2 + \gamma(\phi)\mathcal{G}^2 - {1\over 2}\phi_{,\mu}\phi^{,\mu} - V(\phi)\right]\sqrt{-g}~d^4x,\ee
the Gauss-Bonnet squared term $\mathcal{G}^2$ is essentially of the order of $R^4$, while there is also a $R^2$ term associated with the general scalar-tensor action. This is essentially a generalized version of $F(R, \mathcal{G})$ gravity theory. The functional forms of the coupling parameters are considered for generalization, since in the early universe a scalar field might exist either in the form of a dilaton or Higgs field. Such an action is dealt with recently \cite{I24}, to study the evolution of the very early universe in the background of isotropic and homogeneous Robertson-Walker (RW) metric,

\be\label{RW} ds^2 = -N(t)^2 dt^2 + a(t)^2 \left[ {dr^2\over 1-kr^2} + r^2 d\theta^2 + r^2 \sin^2{\theta} d\phi^2\right],\ee
where, $N(t)$ is the lapse function. The phase-space portrait of the Hamiltonian is then found applying Dirac's constraint analysis. Standard canonically quantization of such higher-order theories suffers from the ambiguities of operator ordering, which is cured to some extent following some ordering prescriptions, such as Weyl ordering \cite{I24.1}. This was done introducing an arbitrary operator ordering parameter $n$. Noticeably, the hermiticity of the effective Hamiltonian demands that $n$ has to be simultaneously equal to $-1$, and $-5$, revealing inconsistency. If one sacrifices Gauss-Bonnet squared term which as mentioned, is essentially of the order of $R^4$, then the Hamiltonian is hermitian for $n = -1$. On the contrary, if one sacrifices $R^2$ term then hermiticity fixes the operator ordering parameter to $n = -5$. Clearly, from physical perspective that a Hamiltonian has to be hermitian, one has to sacrifice either $R^2$ order term or the $R^4$ order Gauss-Bonnet squared term. Since action with $R^2$ term has been extensively studied in the literature, therefore it is worthwhile to study early universe taking into account the latter, sacrificing scalar curvature squared term. However, Gauss-Bonnet term being coupled with a dilatonic field emerges quite naturally as the prime order of the $\alpha'$ expansion of heterotic superstring theory, where, $\alpha'$ is the inverse string tension \cite{I25,I26,I27,I28}. Additionally, the low energy limit of the string theory also induce dilatonic scalar field which couples with different curvature invariant terms \cite{I29,I30}. Therefore the leading quadratic correction produces Gauss-Bonnet term with a dilatonic coupling \cite{I31}. We therefore consider a scalar coupled Gauss-Bonnet term in addition, and the generalized action in the following form,

\be \label{A} A = \int \left[\alpha(\phi)(R - 2\Lambda) +\beta(\phi)\mathcal{G} + \gamma(\phi)\mathcal{G}^2 - {1\over 2}\phi_{,\mu}\phi^{,\mu} - V(\phi)\right]\sqrt{-g}~d^4x,\ee
 is the subject of our present study in connection with the cosmological consequences at the very early as well as the late stages of cosmic evolution. In the context of the early universe, inflation in particular, such study was carried out earlier with a different action, which does not contain Gauss-Bonnet squared term \cite{We2}. In this sense, the present action \eqref{A} is more general. Further, we shall investigate both inflation and cosmic evolution in the radiation-dominated era, in view of the present action, for the first time. It is worth mentioning that although a wonderful feature of gauss-Bonnet-dilatonic coupled term is that, it provides second order field equations and nothing beyond. Therefore it is free from Ostrogradsky's instability; nonetheless, it suffers from the unresolved disease of Branched Hamiltonian \cite{I32,I33,I34,I35,I36,I37,I38,I39}. Attempts to alleviate the pathology following different routes \cite{I35, I36}, led to different canonically inequivalent Hamiltonians \cite{I37}. Thus, canonical formulation of Gauss-Bonnet action remains obscure. Subsequently it was noticed that the pathology due to higher degree terms may be controlled by supplementing the action with higher order ones \cite{I37, I38, I39}. In the above action \eqref{A}, the higher order curvature invariant term appears in the form Gauss-Bonnet squared term, and hence the pathology associated with branched Hamiltonian does not appear, as we shall explore in the present literature. It is noteworthy that, the higher order term (Gauss-Bonnet-squared) invites Ostrogradsky's instability. However, with additional degree of freedom required for canonical formulation, second order field equations appear and as a result the theory becomes free from such instability. We have not considered a term in the form $\mathcal{G}^{n_2}, ~n_2 <{1\over 2}$, to avoid unnecessary complication, since it only becomes dominant at the very late-stage of cosmological evolution. \\

In the above action \eqref{A}, we have intentionally coupled a functional parameter $\gamma(\phi)$ with Gauss-Bonnet squared term for generality. It is further important to mention that, as inflation ends, if the scalar field somehow ceases to evolve at the advent of matter dominated era, then the topologically invariant linear Gauss-Bonnet term will not contribute at the late stage of the evolution of the universe, and would end up with an effective cosmological constant instead of the scalar field, while Gauss-Bonnet squared term might play the role of alternative to the dark energy, as claimed.\\

Usually, an action is studied in the context of the early universe or the late (pressure-less dust era), and sometimes both, to claim unification with early inflation and acceleration at late-stage of cosmic evolution. Here, on the contrary, we shall construct the quantum cosmological equation in the background of RW metric \eqref{RW}, and study its behaviour under an appropriate semiclassical approximation. The reason being, inflation in a quantum theory of perturbation. So only if a theory exhibits a viable semiclassical wave function, then  most of the important physics, such as inflationary dynamics, may be extracted from the classical action. Hence, we study inflation and try to match the inflationary-parameters with latest released results. Finally, we shall examine if the model admits a viable matter dominated (radiation as well as pressure-less dust) era.\\

In section 2, we write down the field equations in the background of homogeneous and isotropic Robertson-Walker space-time \eqref{RW}, and explore de-Sitter solutions. Thereafter in section 3, we prepare the action suitable for canonical formulation, which has finally been formulated applying Dirac's algorithm in Appendix A (canonical quantization, probabilistic interpretation and semiclassical approximations are also performed in Appendix A). The semiclassical approximations exhibits the fact that the semiclassical wave function displays oscillatory behaviour being peaked around the classical de-Sitter solution. This guarantees that the quantum counterpart of the theory under investigation produces a viable classical universe we live in, which allows to study inflation in view of classical field equations. In Appendix B we have computed modified Horowitz' formulation (MHF) to manifest identical canonical structure. We also prove that the effective Hamiltonian is hermitian in Appendix C, and have made some comments regarding the unitarity, which one expects for a physically meaningful theory. Given the complexity of the initial model this is no mean feat. In section 4 we study inflation. Since use of additional hierarchy of the Hubble parameter and the coupling flow parameter (which is a standard procedure to handle additional degrees of freedom) does not work to produce a viable inflation, we have thereafter adopted a technique to reduce extremely complicated field equations to Friedmann-like equations under the choice of three different effective potentials. It is shown that the inflationary parameters are not affected much under such different choices. Finally, we have computed inflationary parameters in Einstein's frame too, which proves consistency of choosing different effective potentials. Section 5 is devoted to study the cosmic evolution in the matter-dominated eras. Concluding remarks appear in section 6.

\section{Field equations and classical de Sitter solutions:}

In the homogeneous and isotropic Robertson-Walker metric \eqref{RW}, the Ricci scalar $R$ and the Gauss-Bonnet term $\mathcal{G}$ are expressed as

\be\label{RG}\begin{split}& R = \frac{6}{N^2}\left(\frac{\ddot a}{a}+\frac{\dot a^2}{a^2}+N^2\frac{k}{a^2}-\frac{\dot N\dot a}{N a}\right) = {6\over N^2}\left[{\ddot z\over 2z} + N^2 {k\over z} - {1\over 2}{\dot N\dot z\over N z}\right]\\&
\mathcal{G} = {24\over N^3 a^3}(N\ddot a - \dot N\dot a)\Big({\dot a^2\over N^2} + k\Big)= {12\over N^2}\left({\ddot z\over z} - {\dot z^2\over 2 z^2} -{\dot N\dot z\over N z}\right)\left({\dot z^2\over 4N^2 z^2} + {k\over z}\right),\end{split}\ee
respectively, where we have translated both, in terms of the induced three metric, $h_{ij} = a^2\delta_{ij} = z\delta_{ij}$, since, as we have noticed earlier \cite{Aks1,Aks2,Aks3,Aks4}, a viable canonical action along with a well-behaved quantum theory may be formulated in terms of the basic variables $\{h_{ij},~K_{ij}\}$, where $K_{ij}$ is the extrinsic curvature tensor. The field equations in connection with the action (\ref{A}) are formulated as ($k = 0$),,

\be\label{zvariation}\begin{split} \\ & 2\alpha\left({\ddot z\over z}-{\dot z^2\over 4z^2} - \Lambda\right)+2\alpha'\left({\ddot\phi}+{\dot\phi}{\dot z\over z} \right)+2\alpha''{\dot\phi^2}+2\left[\frac{\beta''{\dot\phi^2}\dot z^2}{z^2}+\frac{2\beta'{\dot\phi}\dot z\ddot z}{z^2}+\frac{\beta'{\ddot\phi}\dot z^2}{z^2}-\frac{\beta'{\dot\phi}\dot z^3}{z^3}\right]\\&\hspace{0.970 in}+12\gamma\left[{\dot z^4{\ddddot z}\over z^5}+{8\dot z^3{\ddot z}{\dddot z}\over z^5}-{9\dot z^5{\dddot z}\over z^6}+{6\dot z^2\ddot z^3\over z^5}-{135 \dot z^4\ddot z^2 \over 4 z^6}+{159 \dot z^6\ddot z\over 4z^7} - {195 \dot z^8\over 16 z^8}\right]\\&\hspace{.05 in}
+{12\gamma'{\dot\phi}}\left({6\dot z^3 \ddot z^2\over z^5}+{2\dot z^4\dddot z\over z^5}-12{\dot z^5\ddot z\over z^6}+{9\dot z^7\over 2z^7}\right)+{6\gamma''\dot\phi^2}\left({2\dot z^4\ddot z\over z^5}-{\dot z^6\over z^6}\right)+{6\gamma'\ddot\phi}\left({2\dot z^4\ddot z\over z^5}-{\dot z^6\over z^6}\right) \\& \hspace{1.0 in}= -p -\Big[{1\over 2}\dot\phi^2 - V(\phi)\Big].\end{split}\ee
\be\label{00}\begin{split}\\& 2\alpha \left({3\dot z^2\over 4z^2}-\Lambda\right)+{3\alpha'\dot\phi \dot z\over z}+{3\beta'\dot\phi \dot z^3\over z^3} + 18\gamma\left[{\dot z^5 \dddot z\over z^6} + {3\dot z^4 \ddot z^2\over 2z^6} - {9 \dot z^6\ddot z\over 2 z^7} + {15 \dot z^8\over 8 z^8}\right]\\&\hspace{1.10 in}+{18\gamma'\dot\phi \dot z^5\ddot z\over z^6}-{9\gamma'\dot\phi\dot z^7\over z^7}+\frac{81\beta'\gamma'\dot z^{10}}{7z^{10}}= \rho + {1\over 2}\dot\phi^2 + V(\phi).\end{split}\ee
\be\label{phivariation} \begin{split} \\& \ddot\phi + {3\over 2}{\dot z\over z}\dot\phi + V'-{3\alpha'\ddot z\over z}+2\Lambda\alpha'-{3\beta'\dot z^2\ddot z\over z^3}+{{3\beta'\dot z^4}\over 2z^4}-{9\gamma'\dot z^4\ddot z^2\over z^6}-{9\gamma'\dot z^8\over 4z^8}+{9\gamma'\dot z^6\ddot z\over z^7} = 0.\end{split}\ee
In the above, prime denotes derivative with respect to the scalar field $\phi$. The very early universe is vacuum dominated in the absence of perfect or viscous imperfect fluids. In this era ($t \sim 10^{-36} - 10^{-32}$ s.), following some mechanism, as gravity (geometry) enters into the classical domain from the quantum regime via a suitable semiclassical era, the universe is supposed to have entered into an inflationary epoch. Thus, we look for inflationary solution of the above classical field equations setting $\rho = p = 0$, in the following standard de-Sitter form,
\be\label{aphi} a = a_0e^{\lambda t}.\ee
As a result, the above field equations \eqref{00} and \eqref{phivariation} read as,

\be\label{F1}6\alpha\lambda^2+6\alpha'\dot\phi\lambda+24\beta'\dot\phi\lambda^3-576\gamma\lambda^8+1152\gamma'\dot\phi\lambda^7+\frac{82944 \beta'\gamma'\lambda^{10}}{7}-2\alpha\Lambda-V-{{\dot\phi^2}\over{ 2}}=0, \ee
\be \label{F2} \ddot\phi+3\lambda\dot\phi-12\alpha'\lambda^2+2\alpha'\Lambda-24\beta'\lambda^4-576\gamma'\lambda^8+V'=0.\ee
The above pair of field equations \eqref{F1} and \eqref{F2} satisfy the de-Sitter solution \eqref{aphi} provided, the evolution of the scalar field, and the forms of the coupling parameters $\alpha(\phi)$, $\beta(\phi)$, $\gamma(\phi)$, together with the potential $V(\phi)$ are restricted to the following forms:\\

\be\label{param}\begin{split}& \phi= \phi_0e^{-\lambda t};\hspace{0.2 in}\alpha(\phi) = {\alpha_0\over \phi};\hspace{0.2 in}\beta(\phi) = -{\phi^2\over 48\lambda^2}-{\alpha_0\over \phi}\left({1\over 2\lambda^2}-{\Lambda \over 12\lambda^4}\right)=-{{\alpha_0 \beta_0}\over \phi}-{\beta_1 \phi^2};\\&
 V(\phi) = \frac{1}{2} \lambda^2 \phi^2-576\gamma\lambda^8;\hspace{0.2 in} \beta _0=\left({1\over 2\lambda^2}-{\Lambda \over 12\lambda^4}\right);\hspace{0.2 in}\beta_1={1\over 48\lambda^2}; \hspace{0.2 in}\gamma=\gamma_0,\end{split}\ee
where, $a_0$, $\alpha_0$, $\gamma_0$, $\phi_0$, and $\lambda$ are arbitrary constants, while $\beta_0$, $\beta_1$ are not. It is important to mention that, although scalar coupled Gauss-Bonnet models is accountable for late-time cosmic acceleration \cite{2-1, 2-2}, nonetheless, there is absolutely no evidence for the existence of a scalar field in the present universe. Hence, late universe has been probed purely with geometric terms, called the modified theory of gravity. As mentioned, $F(R, \mathcal{G})$ theory, which may contain higher degree Gauss-Bonnet terms, is prevalent amongst all. We deliberately associated a functional form of $\gamma(\phi)$ with $\mathcal{G}^2$ term, and find that de-Sitter solution, which is the foremost requirement for a viable gravity theory, restricts $\gamma = \gamma_0$ - a constant. This is definitely compelling.

\section{Canonical formulation:}

Canonical quantization requires the phase-space structure of the Hamiltonian. We therefore proceed to construct the Hamiltonian corresponding to the action \eqref{A}, in the isotropic and homogeneous minisuperspace \eqref{RW}. It is interesting to note that the classical de-Sitter solution fixed the parameter $\gamma$ to be a constant, which is truly captivating, since higher power of Gauss-Bonnet term was introduced to explain late-time cosmic evolution (acceleration), without seeking help of a scalar field, since no trace of its existence is observed in the present universe. Still, we shall work with $\gamma = \gamma(\phi)$ for generality, and will set it to a constant right in time. The action \eqref{A} in view of the expressions of the scalar curvature and the Gauss-Bonnet term \eqref{RG} may now be expressed as,

\be\begin{split} A& = \int \Bigg[ {6\alpha(\phi)\over N^2}\Big({\ddot z\over 2z} + N^2 {k\over z} - {1\over 2}{\dot N\dot z\over N z}\Big)+\frac{3\beta(\phi)}{N^2z^2}\bigg{(}\frac{{\ddot z}{\dot z^2}}{N^2z}-\frac{\dot z^4}{2N^2 z^2}+\frac{\dot z^3\dot N}{zN^3}+{4k\ddot z}-\frac{2k\dot z^2}{z}-\frac{4k{\dot z}\dot N}{N} \bigg{)}\\& \hspace{0.4 in}+{144\gamma(\phi)\over N^4}\Big({\ddot z\over z} - {\dot z^2\over 2 z^2} -{\dot N\dot z\over N z}\Big)^2\Big({\dot z^2\over 4N^2 z^2} + {k\over z}\Big)^2 +{1\over 2 N^2}\dot \phi^2 - V(\phi) - 2\Lambda \alpha(\phi)\Bigg]Nz^{3\over 2} dt \int d^3 x,\end{split}\ee
and more explicitly to identify the divergent terms as:

\be\begin{split} A = &\int \Bigg[{3\alpha(\phi)\sqrt z\ddot z\over N} + {6\alpha(\phi)\over N}\Big(kN^2 \sqrt{z} - {\dot N\sqrt z\dot z\over 2N}\Big) - 2\alpha(\phi)\Lambda N z^{3\over 2} + N z^{3\over 2}\Big({1\over 2 N^2}\dot\phi^2 - V(\phi)\Big)\\&+\frac{3\beta(\phi)}{N\sqrt z}\bigg{(}\frac{{\ddot z}{\dot z^2}}{N^2z}-\frac{\dot z^4}{2N^2 z^2}+\frac{\dot z^3\dot N}{zN^3}+{4k\ddot z}-\frac{2k\dot z^2}{z}-\frac{4k{\dot z}\dot N}{N} \bigg{)} \\&+ 144\gamma(\phi)\Bigg\{{\ddot z^2\over 16 N^3 z^{9\over 2}}\Big({\dot z^2\over N^2} + 4 k z\Big)^2
-\ddot z\Big({\dot z^6 \over 16 N^7 z^{11\over 2}}+{\dot N\dot z^5\over 8 N^8 z^{9\over 2}}+{k\dot z^4\over 2 N^5z^{9\over 2}}+{k\dot N\dot z^3\over N^6 z^{7\over 2}}+{k^2\dot z^2\over N^3z^{7\over 2}}+{2k^2\dot N\dot z\over N^4 z^{5\over 2}}\Big)\\&
+ {\dot z^8 \over 64 N^7 z^{13\over 2}}+{\dot N\dot z^7\over 16 N^8 z^{11\over 2}} + {\dot N^2\dot z^6\over 16 N^9 z^{9\over 2}}
+{k\dot z^6\over 8 N^5z^{11\over 2}}+{k\dot N\dot z^5\over 2 N^6 z^{9\over 2}} + {k\dot N^2\dot z^4\over 2 N^7 z^{7\over 2}}
+{k^2\dot z^4\over 4N^3z^{9\over 2}}+{k^2\dot N\dot z^3\over N^4 z^{7\over 2}}+{k^2\dot N^2\dot z^2\over N^5 z^{5\over 2}}
\Bigg\}\Bigg]dt.\end{split}\ee
Now under integration by parts divergent terms are removed and the above action is expressed as,

\be\begin{split} \label{A2} &A = \int \Bigg[{6\alpha(\phi) N}\Big(-{\dot z^2\over 4 N^2\sqrt z}+ k  \sqrt z - {\Lambda\over 3} z^{3\over 2}\Big)-\frac{3\alpha'(\phi) \dot\phi \dot z\sqrt z}{N}-\frac{\beta'\dot z\dot\phi}{N\sqrt z}\bigg{(}\frac{\dot z^2}{N^2 z}+12k \bigg{)}\\&+ N z^{3\over 2}\Big({1\over 2 N^2}\dot\phi^2 - V(\phi)\Big) +144\gamma(\phi)\Bigg\{{\ddot z^2\over 16 N^3 z^{9\over 2}}\Big({\dot z^2\over N^2} + 4 k z\Big)^2
- {15\dot z^8 \over 448 N^7 z^{13\over 2}} + {\dot N^2\dot z^6\over 16 N^9 z^{9\over 2}}
-{13 k\dot z^6\over 40 N^5z^{11\over 2}} + {k\dot N^2\dot z^4\over 2 N^7 z^{7\over 2}}\\&
-{11 k^2\dot z^4\over 12 N^3z^{9\over 2}} +{k^2\dot N^2\dot z^2\over N^5 z^{5\over 2}}
-\ddot z\Big({\dot N\dot z^5\over 8 N^8 z^{9\over 2}}+{k\dot N\dot z^3\over N^6 z^{7\over 2}}+{2k^2\dot N\dot z\over N^4 z^{5\over 2}}\Big) +{\gamma'\dot\phi\over \gamma}\Big({\dot z^7\over 112 N^7 z^{11\over 2}} + {k\dot z^5\over 10 N^5 z^{9\over 2}}+ {k^2\dot z^3\over 3 N^3 z^{7\over 2}}\Big)
\Bigg\}\Bigg]dt.\end{split}\ee
It is important to mention that the lapse function $N(t)$, being responsible for diffeomorphic invariance, is supposed to be devoid of dynamics, and should act as a Lagrange multiplier. However, unlike `general theory of relativity', it's very presence in the above action \eqref{A2} with time derivative ($\dot N^2$) might mislead to treat it as a true dynamical variable. Nonetheless, under a change of variable

\be\label{zx} \dot z = N x,\ee
a pair of basic variables are addressed in the form, $h_{ij} = z^2 \delta_{ij}, ~\mathrm{and}~K_{ij} = -{\dot{h_{ij}}\over 2N} = -{a\dot a\over N}\delta_{ij} = -{\dot z\over 2N}\delta_{ij}$, where, $K_{ij}$ is the extrinsic curvature tensor, as mentioned. The action may now be expressed as,

\be\begin{split}\label {A3}& A = \int \Bigg[-{6\alpha(\phi) N}\Big({x^2\over 4\sqrt z} - k\sqrt z + {\Lambda\over 3} z^{3\over 2}\Big)-{3\alpha'(\phi) \dot\phi x\sqrt z} -\frac{\beta'x\dot\phi}{\sqrt z}\bigg{(}\frac{x^2}{z}+12k \bigg{)}+ N z^{3\over 2}\Big({1\over 2 N^2}\dot\phi^2 - V(\phi)\Big)\\&
+ 144\gamma(\phi)\Bigg\{{(x^2 + 4 k z)^2\dot x^2\over 16 N z^{9\over 2}}+{\gamma'(\phi)\dot\phi\over \gamma(\phi)}\Big({x^7\over 112 z^{11\over 2}} + {k x^5\over 10 z^{9\over 2}}+ {k^2 x^3\over 3 z^{7\over 2}}\Big)
- N\Big({15 x^8 \over 448 z^{13\over 2}} + {13 k x^6\over 40 z^{11\over 2}} + {11 k^2 x^4\over 12 z^{9\over 2}}\Big)
\Bigg\}\Bigg]dt.\end{split}\ee
Note that, as a result of introduction of the basic variables ($h_{ij}, ~K_{ij}$), $\dot N$ term disappears from the action, and as such $N$ may be treated just as a Lagrange multiplier. The great triumph of Ostrogradisky's technique \cite{Ostro1, Ostro2} is that, although developed for mechanical problems, it is found to be well-suited to handle higher-order theories of gravity. Now in the above action, both $\dot z$ nor $\dot N$ are absent. Hence, the associated momenta are constrained to vanish, and as a result the Hessian determinant also vanishes. As the resulting point Lagrangian becomes singular, so instead of Ostrogradisky's formalism one should apply Dirac's constrained analysis \cite{Dirac1, Dirac2} (see Appendix A). We also canonically quantized the theory, explored probabilistic interpretation and expatiate a viable semiclassical approximation, to exhibit that the semiclassical wavefunction is oscillatory about the classical de-Sitter solution in Appendix A. This validates the study of slow roll inflation in view of the classical field equations, which we take up in the following section. We have also applied modified Horowitz' formalism (MHF) for the same purpose to demonstrate the fact that identical phase-space structure is manifest (see Appendix B). In Appendix C, we have explored hermiticity of the effective Hamiltonian operator and shed some light on the issue of unitarity.

\section{Slow roll Inflation:}

The recently released data sets \cite{Planck1, Planck2} imposed tighter constraint on $n_s$ ($0.9631 \le n_s \le 0.9705$), as well as on the scalar to tensor ratio ($r < 0.055$). More recently, combination of Planck PR4 data with ground-based experiments such as, BICEP/Keck 2018 (BK18), BAO and CMB lensing data, tightens the scalar to tensor ratio even further to $r < 0.032$ \cite{Planck3}. It is noticeable that in last ten years $r$ has been constrained staring from $r < 0.14$ to the above mentioned value, and therefore we presume that $r$ might be restricted to even less value in future experiments, such as polarized CMB space missions (including LiteBIRD) \cite{Planck4}. Although, all the analysis in regard of such constraints are based on 6-parameter $\Lambda\mathrm{CDM}$ model, nonetheless, it is expected that much more complicated models such as the present one, would only be validated, provided the value of the inflationary parameters do not differ by a large margin. We have exhibited (see Appendix) that following an appropriate semiclassical approximation the quantum universe ($l_P < 10^{-35}$m) transits to the post Planckian era ($l_P > 10^{-35}$m). This is the arena of quantum field theory in curved space-time, where gravity may be treated as classical, while all other fields remain quantized. Therefore inflation, which occurred sometimes between $10^{-42}$s and $10^{-32\pm6}s$, is essentially a quantum mechanical phenomena, specifically, it is a quantum theory of perturbation. However, since the present quantum theory admits a feasible semiclassical approximation, most of the important physics may be extracted from the classical field equations. Let us therefore study inflationary dynamics, find the parameters to compare with currently released data sets \cite{Planck1, Planck2, Planck3}. Therefore, let us rearrange Einstein's ($^0_0$) and the $\phi$ variation equations viz. equations (\ref{00}) and \ref({phivariation}) respectively, for ($\gamma= \gamma_0$ a constant) as,

\begin{center}
\be\label{00H}\begin{split} & \alpha \mathrm{H}^2-\frac{\alpha\Lambda}{3} +\alpha'{\dot\phi}\mathrm{H}+4\beta'{\dot\phi}\mathrm{H}^3+96\gamma_0 \mathrm{H}^8 \Bigg{[}2\bigg\{1+{1\over{ \mathrm{H}^2}}\bigg(\frac{\ddot {\mathrm{H}}}{\mathrm{H}}-2\frac{{\dot {\mathrm{H}}}^2}{\mathrm{H}^2}\bigg)\bigg\}+7\bigg{(}1+\frac{\dot {\mathrm{H}}}{\mathrm{H}^2}\bigg{)}^2\\&\hspace{3.0 in}-8\bigg{(}1+\frac{\dot{\mathrm{H}}}{\mathrm{H}^2}\bigg{)}-2\bigg{]} -\frac{\dot\phi^2}{12}-\frac{V}{6}=0; \end{split} \ee
\end{center}
\be \label{phivar} \begin{split} \ddot\phi +3\mathrm{H}\dot\phi=-V'-2\alpha'\Lambda+&6\alpha'\mathrm{H}^2\bigg{[}\bigg{(}1+\frac{\dot {\mathrm{H}}} {\mathrm{H}^2} \bigg{)}+1\bigg{]}+24\beta'\mathrm{H}^4\bigg{[}\bigg{(}1+\frac{\dot {\mathrm{H}}} {\mathrm{H}^2} \bigg{)}+1\bigg{]}-24\beta'\mathrm{H}^4,
\end{split}\ee
where, $\mathrm{H}={\dot a\over a}$ denotes the slowly varying expansion rate. Note that, inflationary solutions \eqref{param} of the classical field equations (\ref{00}) and (\ref{phivariation}) in standard de-Sitter form, has already been presented . Now, instead of standard slow roll parameters, we introduce a combined hierarchy of the Hubble parameter and the coupling flow parameter, since it is much elegant and convenient \cite{CH1, CH2, CH3, CH4,SR, We1, We2, We3, We4}. In fact, if the background evolution is described by a set of horizon flow functions, it exhibits the nature of Hubble distance during inflation. So we start from,

\be \label{dh}\epsilon_0=\frac{d_{\mathrm{H}}}{d_{\mathrm{H}_i}}, \ee
where $d_{\mathrm{H}}=\mathrm{H}^{-1}$ is the Hubble distance (the horizon) in our chosen units, and the suffix $i$ is used to denote the era at which inflation begins, the hierarchy of functions is defined systematically as:

\begin{center}
    \be \label{el} \epsilon_{l+1}=\frac{d\ln|\epsilon_l|}{d\mathrm{N}},~~l\geq 0, \ee
\end{center}
where, $\mathrm{N}=\ln{\big(\frac{a}{a_i}\big)}$ is the e-fold expansion, and hence $\dot {\mathrm{N}}=\mathrm{H}$. It is now possible to compute the logarithmic change of Hubble distance per e-fold expansion, i.e. $\epsilon_1=\frac{d\ln{d_{\mathrm{H}}}}{d\mathrm{N}}=\dot{d_{\mathrm{H}}}=-\frac{\dot {\mathrm{H} }}{\mathrm{H}^2}$, which is the first slow-roll parameter. This reveals that the Hubble parameter almost remains constant during inflation. From the above hierarchy one can also compute $\epsilon_2=\frac{d\ln{\epsilon_1}}{d\mathrm{N}}=\frac{1}{\mathrm{H}}\big(\frac{\dot\epsilon_1}{\epsilon_1}\big)$, and hence $\epsilon_1\epsilon_2=d_{\mathrm{H}} \ddot{d_{\mathrm{H}}} =-\frac{1}{\mathrm{H}^2}\left(\frac{\ddot {\mathrm{H}}}{\mathrm{H}}-2\frac{\dot {\mathrm{H}}^2}{\mathrm{H}^2}\right)$. Higher slow-roll parameters may similarly be computed. Equation (\ref{el}) described by the equation of motion,

\be\label{el1}\epsilon_0\dot\epsilon_l-\frac{1}{d_{\mathrm{H}_i}}\epsilon_l\epsilon_{l+1}=0,~~~~l\geq 0,\ee
essentially defines a flow in space, where cosmic time acts as the evolution parameter. Using the definition (\ref{dh}), it is also possible to check that equation (\ref{el1}) gives all the results obtained from the hierarchy defined in (\ref{el}). Now, in view of the slow-roll parameters the above equations (\ref{00H}) and (\ref{phivar}) may be expressed as,
 \be \label{hir}\begin{split}& \alpha\mathrm{H}^2-\frac{\alpha\Lambda}{3}+\alpha'\dot\phi\mathrm{H}+4\beta'\dot\phi\mathrm{H}^3+96\gamma_0 \mathrm{H}^8\bigg{[}2\big{(}1-\epsilon_1 \epsilon_2 \big{)}+7\big{(}1-\epsilon_1\big{)}^2-8\big{(}1-\epsilon_1\big{)}-2\bigg{]}-\left(\frac{\dot\phi^2}{12}+{V\over6}\right)=0  \end{split}\ee
and
\be \label{phi}\begin{split}& \ddot\phi+3\mathrm{H}\dot\phi=-V'-2\alpha'\Lambda +6\alpha'\mathrm{H}^2\bigg{[}3-\big{(}1+\epsilon_1\big{)}\bigg{]}+24\beta'\mathrm{H}^4\bigg{[}3-\big{(}1+\epsilon_1\big{)}\bigg{]}-24\beta'\mathrm{H}^4\end{split}, \ee
respectively.
The equations (\ref{hir}) and (\ref{phi}) may therefore be approximated under slow-roll conditions to,

\be\label{hir1} 6\alpha \mathrm{H}^2=576\gamma_0{\mathrm{H}^8}-6\alpha'\dot\phi\mathrm{H}-24\beta'\dot\phi\mathrm{H}^3+\bigg{(}{V}+2\Lambda\alpha\bigg{)}+{{\dot\phi}^2\over 2},\ee
and

\be\label{phi1}\ddot\phi+ 3\mathrm{H}\dot\phi =-V'-2\alpha'\Lambda+12\alpha'\mathrm{H}^2+24\beta'{\mathrm{H}^4}. \ee
 These above set of equations are still formidably complicated to handle, and further simplifications are required. One way is to use additional hierarchy of flow parameters \cite{I38, SR, We1} in connection with additional degrees of freedom $\alpha(\phi)$ and $\beta(\phi)$, associated with the present model. Let us therefore introduce following hierarchy of flow parameters viz.,

\be\label{d1} \delta_1=4\dot \alpha \mathrm{H}\ll 1, ~~~~~\delta_{i+1}=\frac{d\ln|\delta_i|}{d\ln a}, ~~~~\text{with,} ~~~~i\geq 1,\ee

\be\label{eta1} \eta_1=4\dot \beta \mathrm{H}\ll 1, ~~~~~\eta_{i+1}=\frac{d\ln|\eta_i|}{d\ln a}, ~~~~\text{with,} ~~~~i\geq 1.\ee
Clearly, for $i=1, \delta_2=\frac{d\ln|\delta_1|}{dN}=\frac{1}{\delta_1}\frac{\dot\delta_1}{\dot N},$ and $\delta_1\delta_2=\frac{4}{\mathrm{H}}\left(\ddot \alpha \mathrm{H}+\dot \alpha\dot {\mathrm{H}}\right),$ and so on. The slow-roll conditions therefore read as, $|\delta_i| \ll 1$ and $|\eta_i| \ll 1$, which are analogous to the standard slow-roll approximation. In view of the slow-roll parameters, the above equation (\ref{hir1}) may therefore be expressed as,

\be\label{hir2} 6\alpha \mathrm{H}^2=576\gamma_0{\mathrm{H}^8}-{3\over 2}(1+\delta_1)+{3\over 2}-6\mathrm{H}^2(1+\eta_1)+6\mathrm{H}^2+\bigg{(}{V}+2\Lambda\alpha\bigg{)}+{{\dot\phi}^2\over 2} .\ee
Further, using the forms of coupling parameters and $V(\phi)$ \eqref{param}, the equations \eqref{hir2} and \eqref{phi1} reduce to,

\be\label{hcom} 6\alpha \mathrm{H}^2={1\over 2}\lambda^2\phi^2+2\alpha\Lambda+{{\dot\phi}^2\over 2},~~ \text{and}~~~~ \ddot\phi+ 3\mathrm{H}\dot\phi = -2\lambda^2\phi.\ee
Finally, using the standard slow-roll conditions $\dot\phi^2\ll U$ and $|\ddot\phi|\ll 3\mathrm{H}|\dot\phi|$, one obtains,

\be\label{75} 6\alpha \mathrm{H}^2={1\over 2}\lambda^2\phi^2+2\alpha\Lambda,~~ \text{and}~~~~  3\mathrm{H}\dot\phi = -2\lambda^2\phi.\ee
Now, using \eqref{75} one can express the slow-roll parameters as,

\be\label{hirpr}\begin {split}& \epsilon = {\dot{\mathrm{H}}\over \mathrm{H}^2}=\frac{\phi_i^3}{12\alpha_0},~~~~ \eta =2\alpha\left({V''\over V}\right)=\frac{2\alpha_0\lambda^2}{\phi_i\left({1\over 2}\lambda^2\phi^2-576\lambda^8\right)},\\&
\mathrm{N} =\int_{\phi_i}^{\phi_f}{\mathrm{H}\over \dot\phi}d\phi =\int_{\phi_f}^{\phi_i}\left({\phi^2\over 8\alpha_0}+{\Lambda\over 2\lambda^2\phi}\right)d\phi=
{1\over 24\alpha_0}\left(\phi_i^3-\phi_f^3\right)+{\Lambda\over 2\lambda^2}\ln\left({\phi_i\over \phi_f}\right).\end{split}\ee
Unfortunately, these parameters do not render any reasonably good result. The reason is that at the end of inflation $\epsilon_f =1$, the scalar field takes the value $\phi_f^3 ={12\alpha_0}$. Hence, $(\frac{\phi_i}{\phi_f})^3 = \epsilon_i = {r\over 16}$, which is supposed to be very small, resulting in $\phi_i \ll \phi_f$. In some model of-course scalar field increases during inflation, but the present model depicts an exponential decay of the scalar field for de-Sitter solution \eqref{aphi}, hence during slow-roll it can in no way increase. Apart from such inconsistency, note that additional hierarchy oversimplified inflationary parameters \eqref{hirpr} in such a way, that they do not reflect the contribution either from the Gauss-Bonnet square or from Gauss-Bonnet-dilatonic couple terms.\\

Earlier, we followed a completely different technique \cite{We2, We3}. There, we redefined an appropriate potential function, which reduced highly complicated field equations to Friedmann-like equations with a non-minimal coupling, keeping traces of all the additional terms in the redefined potential function. Since, the technique of using additional heirarchy fails, therefore in the underlying three subsections we follow the same route as before \cite{We2, We3} and explore four different cases, under three different choices of the redefined potentials $U(\phi)$. The last case in fact is studied in Einstein's frame to prove consistency of earlier case results.

\subsection{\bf{Case-1:}}

Prior to applying the usual slow-roll conditions, $|\ddot \phi| \ll 3\mathrm{H}|\dot \phi|$ and $\dot\phi^2 \ll V(\phi)$, let us first make attempt to formulate equations \eqref{hir1} and \eqref{phi1} in an unadorned form. Treating $\mathrm{H}$ and $\phi$ as independent variables, we redefine the potential as,

\be\label{U1} U = V +2\alpha\Lambda-12\mathrm{H}^2(\alpha+2\mathrm{H}^2\beta)+576\gamma_0\mathrm{H}^8.\ee
Under such choice, equation (\ref{phi1}) takes the standard form of Klein-Gordon Equation,

\be \label{KG}\begin{split} \ddot\phi +3\mathrm{H}\dot\phi + U'=0.\end{split}\ee
Equation \eqref{KG} implies that like single field equation, the evolution of the scalar field is driven by the so-called re-defined potential gradient $U' = {dU\over d\phi}$, subject to the damping by the Hubble expansion term $3\mathrm{H} \dot \phi$.  The potential $U(\phi)$ carries all the information regarding the coupling parameters of generalised higher order action under current consideration. Further, assuming

\be\label{U2} U = V + 2\alpha\Lambda +576\gamma_0\mathrm{H}^8 - 6\mathrm{H}\dot \phi\left(\alpha' + 4 \mathrm{H}^2 \beta'\right),\ee
equation \eqref{hir1} may also be reduced to the following simplified form, viz,

\be\label{Fried} 6\alpha\mathrm{H}^2 = \frac{\dot\phi^2}{2} + U(\phi).\ee
Importantly, the two choices on the redefined potential $U(\phi)$ made in \eqref{U1} and \eqref{U2}, are at par, since equating the two one obtains,
\be \label{Consistency} \dot\phi (t) = {2\mathrm{H}\alpha + 4\mathrm{H}^3 \beta \over \alpha' + 4 \mathrm{H}^2\beta'},\ee
which on substitution of the forms of coupling parameters and the potential; $\alpha(\phi)$, $\beta(\phi)$, $V(\phi)$ \eqref{param} along with their derivatives, takes the following form,
\be \label{phisol}\dot \phi = {\mathrm{H} \phi\big[4\Lambda \alpha_0 - \mathrm{H}^2 \phi^3\big]\over 4\alpha_0\big(3\mathrm{H}^2 - \Lambda\big) - 2\mathrm{H}^2\phi^3}.\ee
Since the Hubble parameter $\mathrm{H}$ almost remain constant during inflation and is sufficiently small $\mathrm{H}^2 \approx \lambda^2 \approx 10^{-8} M_P^2$, while numerical values that we shall consider to study parameters are $\alpha_0 =0.170 M^3_P,~\Lambda =1 M_P^2$, (see Table 1) one can simply approximate the above equation and express it as,
\be\dot \phi =-{{\lambda \phi(4\alpha_0\Lambda)}\over {4\alpha_0\Lambda}} = -{\lambda \phi}.\ee
Hence $\phi$ decays exponentially with time. The fact that the behaviour of $\phi$ remains unaltered from the de-Sitter solution presented in \eqref{aphi}, is highly appealing, since it validates the choice of the redefined potential. Since consistency of our assumptions is proved, we now apply the standard slow-roll conditions $\dot\phi^2\ll U$ and $|\ddot\phi|\ll 3\mathrm{H}|\dot\phi|$, on equations (\ref{Fried}) and (\ref{KG}), which now take the following forms,

\be\label{H2}{6\alpha}\mathrm{H}^2\simeq U, \ee
and
\be\label{Hphi} 3\mathrm{H}\dot\phi \simeq - U', \ee
respectively. Let us mention that, although under slow- roll approximation, the highly complicated field equations have been reduced to the Friedmann equations, the difference lies in the expression for the redefined potential $U(\phi)$, which contains all the details of higher order terms. Note that the combination of equations (\ref{H2}) and (\ref{Hphi}), exhibits that the potential slow roll parameter $\epsilon$ equals the Hubble slow roll parameter ($\epsilon_1$) under the condition,

\be\label{SR}\begin{split} \epsilon \equiv - {\dot {\mathrm{H}}\over \mathrm{H}^2} = \alpha\left({U'\over U}\right)^2 - \alpha'\left({U'\over U}\right);\hspace{0.3 in}\eta = 2 \alpha \left({U''\over U}\right);\end{split}\ee
while $\eta$ remains unaltered. Additionally, since $\frac{\mathrm{H}}{\dot\phi}=-{U\over2\alpha U'}$, therefore, the number of e-folds at which the present Hubble scale equals the Hubble scale during inflation, can be determined as usual, from the following relation:

\be\label{Nphi} \mathrm{N}(\phi)\simeq \int_{t_i}^{t_f}\mathrm{H}dt=\int_{\phi_i}^{\phi_f}\frac{\mathrm{H}}{\dot\phi}d\phi\simeq \int_{\phi_f}^{\phi_i}\Big{(}\frac{U}{2\alpha U'}\Big{)}d\phi,\ee
where, $\phi_i$ and $\phi_f$ denote the scalar fields at the onset $(t_i)$ and at the end $(t_f)$ of inflation. Thus, slow-roll parameters exhibit all the interactions, as in earlier situations \cite{GB1, GB2, GB3}, but here only through the redefined potential $U(\phi)$.\\

Let us now take up the scalar field potential in the form $V(\phi) = {1\over 2}\lambda^2\phi^2-576\gamma_0\lambda^8$ along with the forms of the coupling parameters $\alpha(\phi)$, $\beta(\phi)$ given in \eqref{param}, which satisfy classical de-Sitter solutions, to compute inflationary parameters numerically. So we need to find the functional form of the re-defined potential $U = U(\phi)$. As mentioned, the Hubble parameter remains almost constant during inflation, and therefore while computing $U(\phi)$, one can replace it by the constant $\lambda$, without much loss of generality. Thus,

\be\label{U} 12\mathrm{H}^2\left(\alpha+2\mathrm{H}^2\beta\right) \approx 2\Lambda\alpha-{{\lambda^2\phi^2}\over 2},~~\mathrm{such~ that},~~ U = V+2\alpha\Lambda+576\gamma_0\lambda^8-2\Lambda\alpha+{{\lambda^2\phi^2}\over 2} ={\lambda^2\phi^2}\approx m^2\phi^2,\ee
where $m$ is the mass of the scalar field. Using the above quadratic form of the re-defined potential \eqref{U}, the expressions for the slow-roll parameters ($\epsilon,~\eta$) \eqref{SR} and the number of e-folds $\mathrm{N}$ \eqref{Nphi} read as,

\be\label{epseta}\epsilon_i = \frac{6\alpha_0}{\phi_i^3},\hspace{0.5 cm}\eta_i = \frac{4\alpha_0}{\phi_i^3}, \hspace{0.5 cm}\mathrm{N} = {1\over 4 \alpha_0}\int_{\phi_f}^{\phi_i} \phi^2 d\phi = {1\over 12\alpha_0}(\phi_i^3 - \phi_f^3),\ee
and one can see that mass ($m$) does not appear. As a result, the utterly complicated theory has been reduced to a two-parameter ($\alpha_0, ~\phi_i$) single field theory. Further, comparing expression for the primordial curvature perturbation on super-Hubble scales produced by single-field inflation ($P_\zeta(k)$) with the primordial gravitational wave power spectrum ($P_t(k)$), the tensor-to-scalar ratio for single-field slow-roll inflation is found as $r = {P_t(k)\over P_\zeta(k)} = 16\epsilon$, while, the scalar tilt, conventionally defined as $n_s-1$ is expressed as $n_s = 1 - 6\epsilon_i + 2\eta_i$. Using all these expressions we now compute the inflationary parameters, which are presented in Table 1, under variation of the parameter $\alpha_0$.\\

\begin{figure}
\begin{center}
   \begin{minipage}[h]{0.47\textwidth}
      \centering
      \begin{tabular}{|c|c|c|c|c|}
     \hline\hline
      $\alpha_0$ in $M_P^3$ & $\phi_f$ in $M_P$ & $n_s$ & $r$ & $\mathrm {N}$\\
      \hline
      0.173 & 1.01251 & 0.9709 & 0.09982 & 80\\
      0.172 & 1.01055 & 0.9711 & 0.09924 & 80\\
      0.171 & 1.00859 & 0.9712 & 0.09867& 81\\
      0.170 & 1.00662 & 0.9714 & 0.09809 &81\\
      0.169 & 1.00465 & 0.9716 & 0.09751 & 81\\
      0.168 & 1.00266 & 0.9717& 0.09694 & 82\\
      0.167 & 1.00067 & 0.9719 & 0.09636 & 82\\
      0.166 & 0.99867 & 0.9721 & 0.09578 & 83\\

      \hline\hline
    \end{tabular}
      \captionof{table}{Data set for the inflationary parameters with $\phi_i=5.5 M_P$, under the variation of $\alpha_0$.}
      \label{tab:table1}
   \end{minipage}%
 \end{center}
\end{figure}

\noindent
Table 1 illustrates that varying $\alpha_0$ within the range $0.173 M_P^3\ge\alpha_0 \ge 0.166 M_P^3$, the spectral index of scalar perturbation and the scalar to tensor ratio remain within the range $0.9709 \le n_s \le 0.9721$ and $0.0998\ge r \ge 0.0957$ respectively while the number of e-folds is found to vary within the range $80 <\mathrm {N} < 83$.  Note that, in the present analysis $n_s$ has already crossed its maximum experimental limit, while the number of e-foldings $\mathrm{N}$ is quite large. Thus, any attempt to lower $r$ value, shifts $n_s$ further beyond the experimental limit, while $\mathrm{N}$ increases even more making the universe supercool at the end of inflation. This will require some additional, currently unknown mechanism to bring the universe back to the hot big bang stage, which is obligatory for the formation of structures and generation of CMB. \\

It can be shown that the so-obtained data sets are independent of the choices of the parameters viz. $\phi_i$ and $\alpha_0$, in the following manner. Let us consider equation \eqref{epseta}, and express $\mathrm{N}$ as,

\be \mathrm{N} ={1\over 12\alpha_0}(\phi_i^3 - \phi_f^3)=\frac{1}{2\epsilon_i}-{1\over 2},\ee
 using the fact that$\phi_f$ is the the value of the scalar field for $\epsilon = 1$, i.e. as inflation ends. Now, setting $r = 0.05$, one finds, $\epsilon_i={0.05\over 16}=.003125$, and, $\eta_i={2\epsilon_i\over 3}= .002083$, in view of \eqref{epseta} again. This leads to $(n_s =0.985)$ and a very large value of $\mathrm{N}\approx 160$, which is independent of $\phi_i$ and $\alpha_0$. On the contrary, if we keep $n_s = 0.97$ at its limiting value, then $\epsilon = 0.00643$, $r = 0.103$ and $N = 77$. Thus, even the requirement that $r < 0.1$ is not fulfilled, keeping other parameters within observational limit. In the following subsection, we therefore redefine the potential in a slightly different manner.

\subsection{\bf{Case-2:}}

Since the previous choice of the potential \eqref{U1} is incompatible with observation, in this case, we therefore redefine the potential as,

\be\label{2U1} U = V +2\alpha\Lambda-12\mathrm{H}^2(\alpha+2\mathrm{H}^2\beta),\ee
excluding the term $576\gamma_0 \mathrm{H}^8$ from \eqref{U1}. The above consideration again modifies equation (\ref{phi1}) to the standard form of Klein-Gordon equation as before,

\be\label{KG2} \ddot\phi +3\mathrm{H}\dot\phi + U'=0.\ee
Further, assuming

\be\label{2U2} U = V + 2\alpha\Lambda +576\gamma_0\mathrm{H}^8 - 6\mathrm{H}\dot \phi\left(\alpha' + 4 \mathrm{H}^2 \beta'\right),\ee
which has the same form as \eqref{U2}, equation \eqref{hir1}  reduces to the following Friedmann-like equation as before, viz.

\be\label{Fried2} 6\alpha\mathrm{H}^2 = \frac{\dot\phi^2}{2} + U(\phi).\ee
Here again note that, the two choices of the redefined potential $U(\phi)$ made in \eqref{2U1} and \eqref{2U2}, do not contradict each other, since equating, one obtains the following first-order differential equation on $\phi$,

\be \label{Consistency2} \dot\phi (t) = {2\mathrm{H}\alpha + 4\mathrm{H}^3 \beta+96\gamma_0\mathrm{H}^7 \over \alpha' + 4 \mathrm{H}^2\beta'} ={\lambda \phi\big[2C_0\phi+4\Lambda \alpha_0 - \lambda^2 \phi^3\big]\over 4\alpha_0\big(3\lambda^2 - \Lambda\big) - 2\lambda^2\phi^3}.\ee
where, we have taken into account the forms of the coupling parameters presented in \eqref{param}, and treated the Hubble parameter to be nearly constant during inflation $\mathrm{H} \approx \lambda$. Now, using the fact that to find the inflationary parameters, the parameters of the theory $C_0$ and $\lambda^2$ have to be $10^{-10}$ order of magnitudes in respective units (see below), which are much smaller than rest of the parameters, the above equation can be suitably approximated to,

\be{\dot\phi (t)}=-{{\lambda \phi(4\alpha_0\Lambda)}\over {4\alpha_0\Lambda}},\hspace{1cm} \Longrightarrow \hspace{1cm} \phi (t)=\phi_ce^{-\lambda t}, \ee
where $\phi_c$ is the constant of integration. The solution for the scalar field displays that even after appreciable (of course admissible) approximations of the field equations to exhibit slow-roll behavior, $\phi$ almost decays exponentially with time, as presented in the de-Sitter solution \eqref{aphi}. As the viability of the chosen pair of effective potentials is proved, we now apply the standard slow-roll conditions $\dot\phi^2\ll U$ and $|\ddot\phi|\ll 3\mathrm{H}|\dot\phi|$, on equations (\ref{Fried2}) and (\ref{KG2}), to obtain,

\be\label{2H2}{6\alpha}\mathrm{H}^2\simeq U, \ee
and
\be\label{2Hphi} 3\mathrm{H}\dot\phi \simeq - U',\ee
respectively. We repeat, as in the previous case, that although under slow roll approximation, the highly complicated field equations have been reduced to the Friedmann equations, the difference lies in the expression for the redefined potential $U(\phi)$, which contains all the information of higher order terms. Let us now compute the functional form of $U = U(\phi)$, as before. For this purpose, let us consider the quadratic form of the potential $V(\phi) = {1\over 2}\lambda^2\phi^2-576\gamma_0\lambda^8$, as in the previous case, along with given forms of $\alpha(\phi)$, $\beta(\phi)$ in \eqref{param}, which satisfy classical de-Sitter solutions. As already mentioned, Hubble parameter remains almost constant during inflation, and hence while computing $U(\phi)$, one can replace it by the constant $\mathrm{H} \approx \lambda$, without losing generality. Thus from \eqref{2U1} one obtains,

\be\label{Uphi} U = {\lambda^2\phi^2}-576\gamma_0\lambda^8={m^2\phi^2}- C_0,\ee
where $m$ may be treated as the mass of the scalar field and $C_0 = 576\gamma_0\lambda^8$. Note that, the only change in the form of $U(\phi)$ from \eqref{U} appears in the form of an additional constant $C_0$. Now, for the above form of $U(\phi)$ \eqref{Uphi}, the inflationary parameters read as,

\be\label{epseta2}\begin{split}&\epsilon_i = \frac{4\alpha_0\phi_i}{(\phi_i^2-{C_0\over m^2})^2}+\frac{2\alpha_0}{(\phi_i^3-\phi_i {C_0\over m^2})},\hspace{0.5 cm}\eta_i = \frac{4\alpha_0}{\phi_i^3-\phi_i {C_0\over m^2}},\\&
\mathrm{N} = {1\over 4\alpha_0}\int_{\phi_f}^{\phi_i}{(\phi^2-{C_0\over m^2})}d\phi = {1\over 12\alpha_0}(\phi_i^3 - \phi_f^3)-{C_0\over 4m^2\alpha_0}(\phi_i-\phi_f).\end{split}\ee
It appears that  we have only been able to reduce the original theory to a four-parameter ($\alpha_0,~ m,~ C_0,~ \phi_i$) one, which is not the characteristics of a good theory. However, we will find that the ratio of $m$ and $C_0$ may be used instead of individuals, thus reducing it to a three-parameter theory. Clearly, the inflationary parameters $\epsilon,~\eta$ and the number of e-folds $\mathrm{N}$ are now modified through the additional parameter $C_0$, which appeared as a constant in the re-defined potential function $U(\phi)$. We now present a pair of data tables varying $\alpha_0$ within the range $0.124 M_P^3\le\alpha_0 \le 0.130 M_P^3$ in Table 2, and varying ${C_0\over m^2}$  within the range $21.4 M_P^2\le {C_0\over m^2} \le 21.9 M_P^2$ in Table 3, keeping $n_s$ within the observational limit, in both. The number of e-folds now varies within the range $58 \le \mathrm{N}\le 61$, in both the tables, which are sufficient to solve the horizon and flatness problems. However, as before, $r > 0.093$ and cannot be reduced any further. In fact, any attempt to keep $r$ within the experimental limit, not only increases $n_s$ beyond observational limit, but also exhibits a spike on the value of number of e-folding. For example, considering the same value of $\phi_i=7.0 M_P$ and choosing $\alpha_0=0.072 M_P^3,~{C_0\over m^2}= 21.5M_P^2$, inflation ends for $\phi_f=4.763 M_P$. In this case, $r = 0.0546$, while $n_s \approx 0.983$, and $\mathrm{N}=105$. Such a huge value of $\mathrm{N}$ does not allow the universe to enter the hot big-bang phase. Further, choosing negative value of $C_0$, which implies the negative coupling of $\mathcal{G}^2$ term, reduces $r$ considerably, but in the process, $N$ increases again.\\
\begin{figure}

   \begin{minipage}[h]{0.47\textwidth}
      \centering
      \begin{tabular}{|c|c|c|c|c|}
      \hline\hline
      $\alpha_0$ in $M^3_P$ & $\phi_f$ in $M_P$ & $n_s$ & $r$ & $\mathrm{N}$\\
      \hline
      0.124& 4.80307 & 0.96988 & 0.09407 & 61\\
      0.125& 4.80375 & 0.96963 & 0.09483 & 60\\
      0.126& 4.80443 & 0.96939 & 0.09559 & 60\\
      0.127& 4.80510 & 0.96915 & 0.09635 & 59\\
      0.128& 4.80578 & 0.96891 & 0.09710 & 59\\
      0.129& 4.80644 & 0.96866 & 0.09786 & 58\\
      0.130& 4.80711 & 0.96842 & 0.09862 & 58\\

       \hline\hline
    \end{tabular}
      \captionof{table}{Data set for the inflationary parameters taking $\phi_i= 7.0M_P$;~~${C_0\over m^2}=21.5 {M^2_P}$ and ~ varying $\alpha_0$, and keeping $n_s$ within Planck constraint limit.}
      \label{tab:table2}
   \end{minipage}%
   \hfill%
  \begin{minipage}[h]{0.47\textwidth}
      \centering
      \begin{tabular}{|c|c|c|c|c|}
      \hline\hline
      ${C_0\over m^2}$ in $ M^2_P$ & $\phi_f$ in $M_P$ & $n_s$ & $r$ & $\mathrm{N}$\\
      \hline
      21.4 & 4.79248 & 0.9701 & 0.09346 & 61\\
      21.5 & 4.80307 & 0.9699 & 0.09407 & 61\\
      21.6 & 4.81364 & 0.9697 & 0.09468 & 60\\
      21.7 & 4.82419 & 0.9695 & 0.09530 & 60\\
      21.8 & 4.83471 & 0.9692 & 0.09593 & 59\\
      21.9 & 4.84521 & 0.9690 & 0.09656 & 59\\
       \hline\hline
    \end{tabular}
      \captionof{table}{Data set for the inflationary parameters taking $\phi_i=7.0 M_P$;~~$\alpha_0=0.124 {M^3_P}$ and ~ varying ${C_0\over m^2}$, and keeping $n_s$ within Planck constraint limit.}
      \label{tab:table3}
   \end{minipage}%
   \end{figure}

It is not possible to explicitly exhibit the fact that the values of inflationary parameters are independent of the choice of the parameters of the theory, as before. However, we take a different route to explore the fact that there is no way to get any better result. From \eqref{epseta2} one can write,

\be\label{97} \epsilon_i =\frac{\eta_i^2\phi_i^3}{4\alpha_0}+\frac{\eta_i}{2}, ~~r =\frac{4\eta_i^2\phi_i^3}{\alpha_0}+{8\eta_i}, ~~ \eta_i={1\over 2}\left(n_s+{3r\over 8}-1\right).\ee
As a result, $\mathrm {N}$ can be expressed as,

\be\label{N98}\mathrm {N}=\frac{4\eta_i-{r\over 8}}{3\eta_i^2} +\frac{2-4\eta_f}{3\eta_f^2}=\frac{2(n_s+{3r\over 8}-1)-{r\over 8}}{{3\over 4}(n_s+{3r\over 8}-1)^2} +\frac{2-4\eta_f}{3\eta_f^2}.\ee
So, from  \eqref{N98}, the desired set of values of $r,~ n_s$ and $\mathrm{N}$, is supposed to give a range of $\eta_f$. Again expressing $\alpha_0$ and $C_0\over m^2$, in terms of $\eta_f$, one can obtain,

\be \alpha_0 = \frac{{\eta_f}^2 {\phi_f}^3}{4-2 {\eta_f}},~~{C_0\over m^2}=\delta =\frac{(3 {\eta_f}-2){\phi_f}^2}{{\eta_f}-2}.\ee
Finally one can also cast $r$ and $n_s$ in terms of $\eta_f$ and $\kappa$, as follows,

\be\label{100}\begin{split}& r =\frac{16 {\eta _f}^2 \kappa ^3 \left[3 {\eta_ f} \left(\kappa ^2-1\right)-2 \kappa ^2+6\right]}{\left[{\eta_f} \left(3 \kappa ^2-1\right)-2 \kappa ^2+2\right]^2},\\&n_s = \frac{{\eta_f}^3 \left(14 \kappa ^3-6 \kappa ^5\right)+{\eta_f}^2 \left(4 \kappa ^5+9 \kappa ^4-28 \kappa ^3-6 \kappa ^2+1\right)-4 {\eta_f} \left(3 \kappa ^4-4 \kappa ^2+1\right)+4 \left(\kappa ^2-1\right)^2}{\left[{\eta_f} \left(3 \kappa ^2-1\right)-2 \kappa ^2+2\right]^2},\end{split}\ee
where $\kappa= {\phi_f \over \phi_i}$. Thus in view of \eqref {N98} and \eqref{100} we observe that, to set $r,~ n_s,~\mathrm {N}$  within experimental limit, one has to fix up a range of $\eta_f$ and $\kappa$. The range is graphically demonstrated in Fig-1, where $\kappa$ has been plotted along abscissa, and  $\eta_f$ along the ordinate. In Figure-1, the blue, the red and the green dots correspond to the ranges of $r$ between $0.05 < r < 0.08$, $n_s$ between $0.960 < n_s < 0.975$ and $\mathrm{N}$ within $50 <\mathrm{N} < 72$, respectively. Thus the overlapping region of blue, green and red which is depicted by the region \emph{AB} corresponds to the values of the inflationary parameters $r, ~n_s $ and $\mathrm{N}$, simultaneously. We peak a few data points $n_f$ and $\kappa$ from the overlapping region \emph{AB}, and compute the inflationary parameters, which are exhibited in Table-4. Clearly, attempt to reduce $r$ tells upon $n_s$ and $\mathrm{N}$, and in no way we can set the inflationary parameters within the observational limit.\\

\begin{figure}
\begin{minipage}[h]{0.47\textwidth}
\centering
\includegraphics[ width=1.35\textwidth] {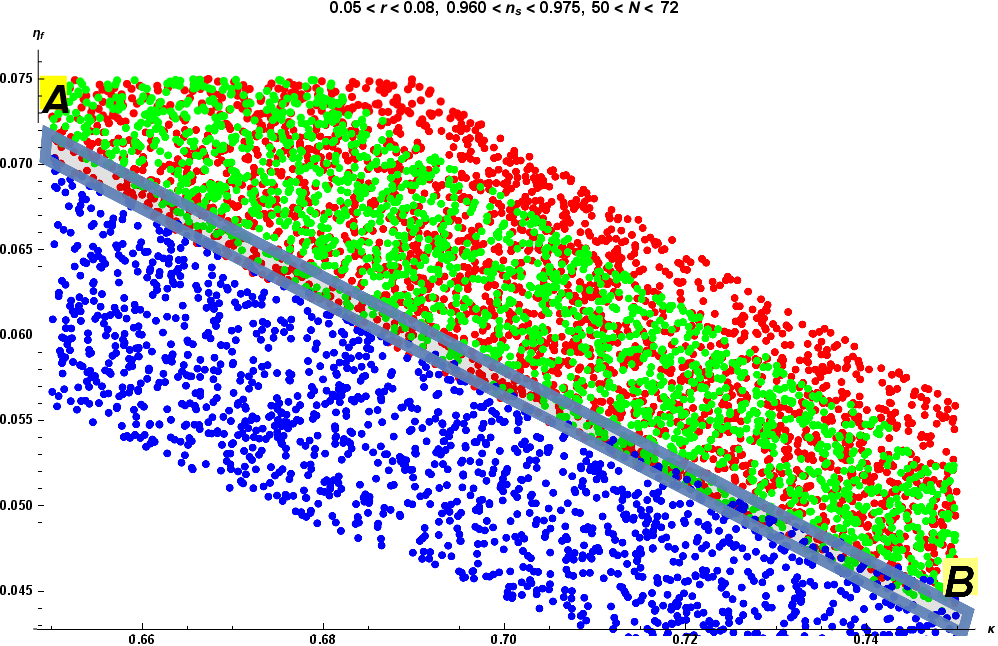}
 \caption{This plot depicts ranges of $\kappa$ and $\eta_f$ .}
      \label{fig:1}
   \end{minipage}%
  \hfill
\begin{minipage}[h]{0.35\textwidth}
\centering
\begin{tabular}{|c|c|c|c|c|}
      \hline\hline
      $\eta _f$ & $\kappa$ & $n_s$ & $r$ & $\mathrm{N}$\\
      \hline
      0.04611 & 0.7401 & 0.9750 & 0.07652 & 71\\
      0.04743 & 0.7358 & 0.9746 & 0.07761 & 70\\
      0.04848 & 0.7320 & 0.9745 & 0.07815 & 70\\
      0.04892 & 0.7296 & 0.9747 & 0.07778 & 70\\
      0.05007 & 0.7257 & 0.9745 & 0.07850 & 70\\
      0.05103 & 0.7220 & 0.9744 & 0.07875 & 70\\
      0.05174 & 0.7194 & 0.9744 & 0.07900 & 70\\
      0.05290 & 0.7149 & 0.9744 & 0.07920 & 70\\
      0.05385 & 0.7116 & 0.9743 & 0.07961 & 70\\
      0.05710 & 0.6996 & 0.9742 & 0.08026 & 70\\
      \hline\hline
    \end{tabular}
      \captionof{table}{Data set for the inflationary parameters extracting $\eta_f$ and $\kappa$ from the common region as depicted in Fig-1.}
      \label{tab:table4}
   \end{minipage}%
\end{figure}

Nonetheless, the present model behaves appreciably otherwise, as it exhibits post-Planckian energy scale of inflation, and also admits graceful exit. For example, taking into account $U(\phi)$ \eqref{Uphi}, we find the following expression from equation \eqref{Fried2},

\be \label{HM}6{\alpha_0\over\phi}\mathrm{H}^2= m^2(\phi^2-{C_0\over m^2}).\ee
Now, if we choose a mid range value of ${C_0\over m^2} = 21.5 M^2_P$, together with a mid range value of $\alpha_0 =0.127 M^3_P$,  with, ~$\phi_i=7.0 M_P$, as depicted in the table 2 and table 3, we simply find,

\be \mathrm{H}^2=\frac{m^2(\phi^3-{C_0\phi\over m^2})}{6\alpha_0}, \hspace{0.3in} \mathrm{and~hence}, \hspace{0.3in}  \mathrm{H} ^2 \approx 253m^2,\ee
showing consistency with our data set displayed in Table 2 and Table 3. So, the energy scale of inflation$(\mathrm{H_*}\approx 15.9m)$ is sub-Planckian, since as mentioned earlier,$m \approx 10^{-5} M_P$. Further, since $r \approx 0.09$ for the above choice of $\alpha_0$, so we can easily calculate the energy scale of inflation $\mathrm{H_*}$ using the formula for single scalar field model as \cite{Wands},

\be \mathrm{H_*}=8\times10^{13}{\sqrt{r\over 0.2}}GeV = 5.36 \times 10^{13} GeV \approx 2.2\times 10^{-5}M_P.\ee
$\mathrm H_*$ in the present case is just one order of magnitude higher than the single field inflation, which is the outcome of the redefined potential containing all the information regarding higher order terms and interactions. \\

As mentioned the model gracefully exits from inflation as well, since as $\phi \ll 1 M_P$, it executes oscillatory behaviour, as shown below. Let us now express equation \eqref{Fried2} as,

\be 3\mathrm{H}^2 = {1\over 2\alpha}\left({1\over 2}\dot\phi^2 + m^2 \phi^2-C_0\right),\ee
taking, $U(\phi) =  m^2 \phi^2-C_0$. In view of the expression of $\alpha(\phi) = {\alpha_0\over \phi}$, the above equation reads as,

\be \label{Hubble} {3\mathrm{H}^2\over m^2} = {\phi\over 2\alpha_0}\left({\dot\phi^2\over 2m^2} + \phi^2-{C_0\over m^2}\right).\ee
Note that, for GTR associated with non-minimally coupled single scalar field model, the above equation simplifies to: $3\mathrm{H}^2 = {1\over 2M_p^2}(\dot\phi^2 + 2 m^2 \phi^2-2C_0)$. Since as inflation ends, ${\phi\over 2\alpha_0} \sim {28 M_p^{-2}}$ according to the present data set, so as the Hubble rate ($\mathrm{H}$) falls below $m$, this equation \eqref{Hubble} can be approximated to,

\be \dot\phi^2 \approx -2(m^2 \phi^2- C_0),\ee
which may be integrated to yield,

\be\phi (t)=\pm \frac{\sqrt{C_0} \tan \left[m (\sqrt{2} t-t_0)  \right]}{m  \sqrt{\tan ^2\left[ m  (\sqrt{2}t-t_0) \right]+1}},\ee
and may further be simplified to obtain
\be\phi (t)=\pm \frac{\sqrt{C_0}}{m}\sin \left[m (\sqrt{2} t-t_0)\right]. \ee
Where $t_0$ is the constant of integration. Thus the scalar field starts oscillating many times over a Hubble time, driving a matter-dominated era when inflation terminates.

\subsection{\bf{Case-3:}}

Since, other than the value of $r$, the model behaves appreciably, it needs further study. In this subsection, we consider the last and the most legitimate (the reason will be clear shortly) choice of the effective potential as,

\be\label{3U1} U = V +2\alpha\Lambda + 6\alpha{\dot{\mathrm{H}}}-24\beta\mathrm{H}^4.\ee
A comparison with \eqref{2U1} reveals that, the present potential \eqref{3U1} is cast excluding the term $-12 \alpha \mathrm{H}^2$ and including $6\alpha \dot{\mathrm{H}}$,instead. This choice (\ref{3U1}) again leads to klein-Gordon equation \eqref{phi1}. Further, if we choose

\be\label{3U2} U = V +2\alpha\Lambda-24\beta'{\dot\phi}\mathrm{H}^3+576\gamma_0 \mathrm{H}^8, \ee
excluding$-6\alpha'{\dot\phi}\mathrm{H}$ term from \eqref{2U2}, Friedman equation \eqref{hir1} may also be retrieved. Now, from \eqref{3U1}, one can express $U = U(\phi)$ as,

\be\label{U3} U =\lambda^2\phi^2-576\gamma_0 {\lambda}^8+\frac{12\alpha_0{\lambda}^2}{\phi}= m^2\phi^2+\frac{12\alpha_0 m^2}{\phi}-C_0,\ee
where the solutions of $V, ~\beta$ are substituted from equation \eqref{param}, with the choice ${\lambda}^2= m^2,~576\gamma_0 {\lambda}^8 =C_0$ and neglecting the term $6\alpha{\dot{\mathrm{H}}}$, since ${\dot{\mathrm{H}}}$ is nearly constant during inflation. The reason for considering the effective potentials in the above forms \eqref{3U1}, and \eqref{3U2}, is unveiled from the following scalar-tensor theoretic action of gravity,

\be \label{A3a} A = \int \left[\alpha(\phi)R  - {1\over 2}\phi_{,\mu}\phi^{,\mu} - U(\phi)\right]\sqrt{-g}~d^4x.\ee
The above action \eqref{A3a} leads to the following field equations in the R-W space-time \eqref{RW} under present consideration,

\be \label{FE3} 6\alpha{\dot a^2\over a^2}= -6\alpha'{\dot a\over a}\dot\phi+U+{1\over 2}{\dot\phi}^2; \hspace{1.0cm}{\ddot\phi} +3{\dot a\over a}\dot\phi = -U'+6\alpha'{\left({\ddot a\over a}-{\dot a^2\over a^2}\right)}+12\alpha'{\dot a^2\over a^2}.\ee
The field equations \eqref{hir1} and \eqref{phi1} are retrieved from the above pair under substitution of the expression of $U$ chosen in \eqref{3U2} and \eqref{3U1} respectively. The theory has thus been reduced to a scalar-tensor theory of gravity. Now, there are two independent ways to study inflation. The first is to proceed as before in the Jordan frame, and the second is to translate the action in Einstein frame and proceed. We study these two cases in the following two subsections.

\subsubsection{Inflation in Jordan frame:}

In view of the expression for $U(\phi)$ \eqref{U3} the following inflationary parameters are found:
\be \label{epeta3}\begin{split}\epsilon& = {\alpha_0\over \phi}\left(\frac{2\phi-{12\alpha_0\over \phi^2}}{\phi^2+{12\alpha_0\over \phi}-{C_0\over m^2}}\right)^2+{\alpha_0\over \phi^2}\left(\frac{2\phi-{12\alpha_0\over \phi^2}}{\phi^2+{12\alpha_0\over \phi}-{C_0\over m^2}}\right);\\&\eta= {2\alpha_0\over \phi}\left(\frac{2+{24\alpha_0\over \phi^3}}{\phi^2+{12\alpha_0\over \phi}-{C_0\over m^2}}\right);\hspace{2.0 cm}
 \mathrm{N}={1\over 2\alpha_0}\int_{\phi_f}^{\phi_i}{\phi(\phi^2+{12\alpha_0\over \phi}-{C_0\over m^2})\over{(2\phi-{12\alpha_0\over \phi^2})} }d\phi.\end{split}\ee
\begin{figure}
\begin{minipage}[h]{0.47\textwidth}
      \centering
      \begin{tabular}{|c|c|c|c|c|}
      \hline\hline
      ${C_0\over m^2}$ in $M^2_P$ & $\phi_f$ in $M_P$ & $n_s$ & $r$ & $\mathrm{N}$\\
      \hline
      144.18& 0.84318 & 0.96502 & 0.01995 & 47\\
      144.20& 0.84314 & 0.96504 & 0.01991 & 51\\
      144.22& 0.84310 & 0.96506& 0.01988 & 54\\
      144.24& 0.84306 & 0.96508 & 0.01985 & 57\\
      144.26& 0.84302 & 0.96510 & 0.01981 & 59\\
      144.28& 0.84298 & 0.96511 & 0.01978 & 62\\
       \hline\hline
    \end{tabular}
      \captionof{table}{Data set for the inflationary parameters taking $\phi_i=7.26 M_P$;~$\alpha_0 = 2.0 {M^3_P}$}
      \label{tab:table}
   \end{minipage}%
 \hfill
\begin{minipage}[h]{0.47\textwidth}
\centering
\begin{tabular}{|c|c|c|c|c|}
      \hline\hline
      $\alpha_0$ in $M^3_P$ & $\phi_f$ in $M_P$ & $n_s$ & $r$ & $\mathrm{N}$\\
      \hline
      0.0074&6.96004  &0.9707  &0.08113  &53 \\
      0.0075&6.96025  &0.9703  &0.08222  &53 \\
      0.0076& 6.96045 &0.9699  &0.08331  &52 \\
      0.0077&6.96066  &0.9696  &0.08440  &51 \\
      0.0078&6.96086  &0.9691  &0.08550  &51 \\
      0.0079&6.96107  &0.9688  &0.08659  &50 \\
      0.0080&6.96127  &0.9684  &0.08768  &49 \\
       \hline\hline
    \end{tabular}
      \captionof{table}{Data set for the inflationary parameters taking $\phi_i=7.40 M_P$;~${C_0\over m^2} = 48 {M^2_P}$}
      \label{tab:table}
   \end{minipage}%

      \end{figure}
A pair of data sets are exhibited in Table-5 and Table-6. Table-5, in which ${C_0\over m^2}$ is varied in the range $144.18 M_P^2 <{C_0\over m^2} < 144.28 M_P^2$, keeping $\phi_i = 7.26 M_P$ and $\alpha_0 = 2.0 M_P^3$, shows tremendous fit with observation. On the contrary, Table-6, in which ${\alpha_0}$ is varied in the range $.0074 M_P^3<{\alpha_0} < .0080 M_P^3$, keeping $\phi_i = 7.40 M_P$, and ${C_0\over m^2} = 48 M_P^2$, shows nothing better than previous results obtained in earlier cases (Table-1 to Table-4). Note that in Table-6, $n_s$ is kept within the observational limit.\\

Since in earlier cases we have shown that the results are independent of the choice of the parameters of the theory the excellent fit in Table-5 therefore develops curiosity, and requires further study. Let us therefore proceed to find the energy scale of inflation. Using the above form of $U(\phi)$ \eqref{U3}, we procure the following expression from equation \eqref{Fried2},

\be 6{\alpha_0\over\phi}\mathrm{H}^2= m^2\phi^2 + {12\alpha_0 m^2\over \phi}-C_0, ~~\mathrm{or,}~~ \mathrm{H}^2 =\frac{6\alpha_0}{m^2}\left(\phi^3 + 12\alpha_0-{C_0\phi\over m^2}\right).\ee
Taking $\phi_i=7.26 M_P,~{C_0\over m^2}=144.22 M_P^2,~\alpha_0= 2M_P^3,$ from table-5 we find,
\be\mathrm {H}^2= -53.37 m^2M_P^2,\ee
Unfortunately, the wonderful fit of the present theory with observational data appears at a high price, viz. imaginary value of the energy scale, and therefore the data of Table-5 should be disregarded. On the contrary, the data presented in Table-6, viz. $\phi_i=7.40M_P,~{C_0\over m^2}=48 M_P^2,~\alpha_0= .008 M_P^3$ yields $\mathrm {H}^2= 522 m^2 M_P^2$, and so the energy scale of inflation is $\mathrm{H_*}=2.3\times 10^{-4}M_P$, taking $m^2\approx 10^{-10}M_P^2$, which is again sub-Planckian.

\subsubsection{Inflation in Einstein frame:}

So far our study in all the above cases reveals that in no way, the tensor to scalar ratio may be reduced and matched with recent observation, keeping the scalar tilt within the observational limit ($0.9631 \le n_s \le 0.9705$), and the number of e-folds within reasonable range $45 < \mathrm{N} < 70$. Finally, in the current subsection we study inflation in Einstein's frame to explore better result in any. The Jordan frame action \eqref{A3} can be translated to the Einstein-frame under the conformal transformation, $g_{E\mu\nu} = f(\phi) g_{\mu\nu}$ \cite{Conformal},\cite{Beh} as,

\be\label{A4}
A_E = \int \left[R_E  - {1\over 2}\sigma_{E,\mu}\sigma^{E,\mu} - U_E(\sigma)\right]\sqrt{-g_E}~d^4x,\ee
where, the subscript $`E$' denotes Einstein's frame. The above action \eqref{A3} leads to the following field equations,

\be\label{FE4} 3\mathrm {H}^2=U_E+{1\over 2}{\dot\sigma}^2; \hspace{1cm}\mathrm{and}\hspace{1cm}{\ddot\sigma} +3\mathrm {H}{\dot\sigma} = -U'_E,\ee
where $U_E ={U(\phi)\over{\alpha(\phi)}^2}$ and $(\sigma')^2= \left(\frac{d\sigma}{d\phi}\right)^{2}= \frac{1}{\alpha(\phi)}+3\frac{\alpha'(\phi)^2}{\alpha(\phi)^2}$.
In the present case $\alpha(\phi)={\alpha_0\over \phi}$ and $(\sigma')^2=\frac{\phi}{\alpha_0}+\frac{3}{\phi^2}$, and $U(\phi)= m^2\phi^2+\frac{12m^2\alpha_0}{\phi}-C_0$. The slow roll parameters can be written as shown in \cite{Beh}, in the following forms,

\be\epsilon= \left(\frac{U'_E}{U_E}\right)^2\left(\frac{d\sigma}{d\phi}\right)^{-2}=\left[\frac{4\phi^3+12\alpha_0-2{C_0\phi\over m^2}}{\phi^4+12\alpha_0\phi-{C_0\phi^2\over m^2}}\right]^2{1\over ({\phi\over \alpha_0}+{3\over \phi^2})},\ee
\be\begin{split}\eta&=2\left[{\frac{U''_E}{U_E}}\left(\frac{d\sigma}{d\phi}\right)^{-2}-
{\frac{U'_E}{U_E}}\left(\frac{d\sigma}{d\phi}\right)^{-3}\frac{d^2\sigma}{d\phi^2}\right] \\&=2\left[\frac{12\phi^2-2{C_0\over m^2}}{\left( \phi^2+\frac{12\alpha_0}{\phi}-{C_0\over m^2} \right)({\phi^3\over \alpha_0}+3)}-\frac{\left(4\phi^3+12\alpha_0-2{C_0\over m^2}\phi\right)\left({1\over \alpha_0}-{6\over \phi^3}\right)}{2\phi^2\left(\phi^2+\frac{12\alpha_0}{\phi}-{C_0\over m^2}\right)({\phi\over \alpha_0}+{3\over \phi^2})^2}\right].\end{split}\ee
\be \mathrm{N}={1\over 2}\int {d\phi\over \epsilon(\phi)} {d\sigma\over d\phi}={1\over 2}\int\left({\phi\over \alpha_0}+{3\over \phi^2}\right)\left(\frac{\phi^4+12\alpha_0\phi-{C_0\phi^2\over m^2}}{4\phi^3+12\alpha_0-2{C_0\phi\over m^2}}\right){d\phi}.\ee

\begin{figure}
\begin{minipage}[h]{0.47\textwidth}
      \centering
      \begin{tabular}{|c|c|c|c|c|}
      \hline\hline
      ${\alpha_0}$ in $M^2_P$ & $\phi_f$ in $M_P$ & $n_s$ & $r$ & $\mathrm{N}$\\
      \hline
      0.0061& 7.0999 & 0.9701 & 0.09226 & 51\\
      0.0062& 7.1002 & 0.9697 & 0.09377 & 50\\
      0.0063& 7.1004& 0.9692 & 0.09528 & 49\\
      0.0064& 7.1006 & 0.9687 & 0.09678 & 48\\
      0.0065& 7.1009 & 0.9682 & 0.09829 & 47\\
      0.0066& 7.1011 & 0.9676 & 0.09980 & 46\\
       \hline\hline
    \end{tabular}
      \captionof{table}{Data set for the inflationary parameters taking $\phi_i=7.5 M_P$;~${C_0\over m^2} = 50 {M^2_P}$}
      \label{tab:table}
   \end{minipage}%
   \hfill
\begin{minipage}[h]{0.47\textwidth}
\centering
\begin{tabular}{|c|c|c|c|c|}
      \hline\hline
${C_0\over m^2}$ in $M^2_P$ & $\phi_f$ in $M_P$ & $n_s$ & $r$ & $\mathrm{N}$\\
      \hline
      140& 0.92287 & 0.96114 & 0.02563 & 65\\
      142& 0.92213 & 0.96145 & 0.02698 & 64\\
      144& 0.92141 & 0.96173 & 0.02825 & 64\\
      146& 0.92071 & 0.96199 & 0.02952 & 63\\
      148& 0.92003 & 0.96224 & 0.03077 & 62\\
      150& 0.91937 & 0.96246 & 0.03199 & 62\\
      \hline\hline
    \end{tabular}
      \captionof{table}{Data set for the inflationary parameters taking $\phi_i=7.0 M_P$;~$\alpha_0 = 0.65 {M^3_P}$}
      \label{tab:table}
   \end{minipage}%
\end{figure}
\noindent
Here we present two tables, viz. Table-7 and Table-8. It appears that Table-8 fits observational data with extreme precession. Let us, therefore explore the energy scale of inflation for both the cases. Considering the above form of $U_E(\phi)$ \eqref{3U1}, we obtain the following expression from equation \eqref{Fried2},

\be 3\mathrm{H}^2= (m^2\phi^2+\frac{12m^2\alpha_0}{\phi}-C_0)= m^2(\phi^2+\frac{12\alpha_0}{\phi}-{C_0\over m^2}).\ee
Now, taking $\phi_i=7.5M_P,~{C_0\over m^2}=50 M_P^2,~\alpha_0= 0.0064M_P^3,$ from Table-7 one can get,
\be3\mathrm{H}^2= {(56.25+0.01024-50)m^2}, ~~~or,~~\mathrm{H}^2=2.08\times 10^{-10}M_P^2,\ee
with $m^2\approx 10^{-10}M_P^2$. So, the energy scale of inflation is $( \mathrm{H_*}\approx1.44\times 10^{-5}M_P)$ is sub-Planckian and fits with single field energy scale perfectly. On the contrary, taking $\phi_i=7.0M_p,~{C_0\over m^2}=146 M_P^2,~\alpha_0= 0.65M_P^3,$ from Table-8 one obtains,

\be\mathrm {H}^2= \frac{(49+1.11-146)m^2}{3}, ~~~ \mathrm{H_*}^2=-32\times 10^{-10}M^2_P,\ee
with $m^2\approx 10^{-10}M_P^2$. So, the energy scale of inflation is negative, and hence data in Table-8 are faulty. In a nut-shell, present analysis exhibits that the inflationary parameters remain unaffected under different choices of the redefined potentials, together with the one in Einstein's frame. It has also been shown that the inflationary parameters are independent of the choice of the parameters of the theory.

\section{Cosmic evolution in the matter dominated era:}

At this end we observe that, the gravitational action for non-minimally coupled scalar-tensor theory, being supplemented by a Gauss-Bonnet term with scalar coupling and a Gauss-Bonnet squared term, leads to a well-behaved quantum picture (The effective Hamiltonian is hermitian, standard quantum mechanical probabilistic interpretation holds, and transition over to the classical universe under a viable semi-classical approximation is admissible). However, the action is not quite consistent with recent observational constraints on inflationary parameters. It's true that the action is well suited to act as an alternative to dark energy at the very late stage of cosmic evolution. It is therefore necessary to study cosmic evolution in the middle - the radiation dominated era which initiated soon after graceful exit from inflation, and the early pressure-less dust era. The oscillation of the scalar field produces particles which heats up the universe whence it enters the so called `hot big bang' era, which is the onset of radiation dominated era. All the essential features at this era (structure formation and the formation of CMBR right in time) are perfectly described by the Friedmann-solution ($a =a_0\sqrt t$). With the evolution, the universe reaches a stage called `matter-radiation equality' at a redshift around $z \sim 3200$. Thereafter matter takes over radiation and the universe enters a pressureless dust era. This era again is best described by Friedmann solution ($a\propto t^{2\over 3}$) until recently ($z \sim 1$), whence the universe evolves with accelerated expansion. Note that, the de-Sitter solution has already fixed the forms of the coupling parameters $\alpha(\phi) = {\alpha_0\over \phi},~\beta(\phi) = -{\alpha_0\beta_0\over \phi} - \beta_1\phi^2$, and $\gamma(\phi) = \gamma_0$ = constant, along with the potential $V(\phi) = m^2\phi^2 - V_0$ \eqref{param}, and in no way one can consider else.\\

To study the evolutions in the matter-dominated (radiation and early pressureless dust) eras, we therefore write the field equations \eqref{zvariation}, \eqref{00} and \eqref{phivariation} in flat space $k = 0$, including a barotropic fluid, and substituting the forms of $\alpha(\phi),~\beta(\phi),~\gamma(\phi) = \gamma_0$ etc., along with potential $V(\phi)$ obtained for de-Sitter solution in \eqref{param} as,

\be\begin{split} \label{fe1} & 2{\alpha_0\over \phi}\left({\ddot z\over z}-{\dot z^2\over 4z^2} - \Lambda\right)-2{\alpha_0\over \phi^2}\left({\ddot\phi}+{\dot\phi}{\dot z\over z} \right)+4{\alpha_0\over \phi^3}{\dot\phi^2}-\frac{{\dot\phi^2}\dot z^2}{z^2}\left[{1\over 12\lambda^2}+{2\alpha_0 \over \phi^3}\left(\frac{1}{\lambda^2}-\frac{\Lambda}{6\lambda^4}\right)\right]\\&\hspace{0.3 in}+\left\{\frac{2{\dot\phi}\dot z\ddot z}{z^2}\left[-{\phi\over 12\lambda^2}+{\alpha_0 \over \phi^2}\left(\frac{1}{\lambda^2}-\frac{\Lambda}{6\lambda^4}\right)\right]+\frac{{\ddot\phi}\dot z^2}{z^2}\left[-{\phi\over 12\lambda^2}+{\alpha_0 \over \phi^2}\left(\frac{1}{\lambda^2}-\frac{\Lambda}{6\lambda^4}\right)\right]\right\}\\&\hspace{0.3 in}+12\gamma_0\left[{\dot z^4{\ddddot z}\over z^5}+{8\dot z^3\ddot z\dddot z\over z^5}-{9\dot z^5\dddot z\over z^6}+{6\dot z^2\ddot z^3\over z^5}-{135 \dot z^4\ddot z^2 \over 4 z^6}+{159 \dot z^6\ddot z\over 4z^7} - {195 \dot z^8\over 16 z^8}\right] \\&\hspace{0.3 in} -\frac{{\dot\phi}\dot z^3}{z^3}\left[-{\phi\over 12\lambda^2}+{\alpha_0 \over \phi^2}\left(\frac{1}{\lambda^2}-\frac{\Lambda}{6\lambda^4}\right)\right]= -\omega{\rho_0}{a^{-3(1+\omega)}} -\Big[{1\over 2}\dot\phi^2 -\left({\lambda^2\phi^2\over 2}-C_0\right)\Big];\\&
2{\alpha_0\over \phi} \left({3\dot z^2\over 4z^2}-\Lambda\right)-{3\alpha_0\dot\phi \dot z\over \phi z}+{3\dot\phi \dot z^3\over z^3}\left[-{\phi\over 24\lambda^2}+{\alpha_0 \over \phi^2}\left(\frac{1}{2\lambda^2}-\frac{\Lambda}{12\lambda^4}\right)\right] \\&
\hspace{0.3 in}+ 18\gamma_0\left[{\dot z^5 \dddot z\over z^6} + {3\dot z^4 \ddot z^2\over 2z^6} - {9 \dot z^6\ddot z\over 2 z^7} + {15 \dot z^8\over 8 z^8}\right] = {\rho_0}{a^{-3(1+\omega)}} + {1\over 2}\dot\phi^2 +{\lambda^2\phi^2\over 2}-C_0 ;\\&
\ddot\phi + {3\over 2}{\dot z\over z}\dot\phi+{3\alpha_0\ddot z\over z \phi^2}-2\Lambda{\alpha_0\over \phi^2}-{3\dot z^2\ddot z\over z^3}\left[-{\phi\over 24\lambda^2}+{\alpha_0 \over \phi^2}\left(\frac{1}{2\lambda^2}-\frac{\Lambda}{12\lambda^4}\right)\right]\\&\hspace{0.3 in}+{{3\dot z^4}\over 2z^4}\left[-{\phi\over 24\lambda^2}+{\alpha_0 \over \phi^2}\left(\frac{1}{2\lambda^2}-\frac{\Lambda}{12\lambda^4}\right)\right] +\lambda^2\phi = 0.\end{split}\ee
Let us now seek a power law solution $a \propto t^{n}$ i.e. $z = a^2 = c^2(t)^{2n}$ (say) and $\phi = \phi_0t^{-m}$(say), using the barotropic equation of state, $p = {\omega\rho}$, where, $\omega={1\over3}$ for radiation dominated era and $\omega=0$ for pressureless dust era. The Bianchi identity $\dot\rho + 3{\dot a\over a}(\rho + \omega\rho) = 0$ therefore results in,

\be\label{bi} \rho = \rho_0 a^{-3(1+\omega)},\ee
where, $\rho_0$ represents the value of the density (radiation or matter) available in the present universe. The above set of field equations \eqref{fe1} are therefore expressed as:

\be\label{MZ}\begin{split}& {2 \alpha_0  \over \phi_0t^{-m}}\left({2n(2n-1)\over t^2}-{n^2\over t^2}-\Lambda\right)-{2 \alpha_0  \over \phi_0^2t^{-2m}}\left({\phi_0m(m+1)\over t^(m+2)}-{2nm\phi_0\over t^{m+2}}\right)\\&\hspace{0.3 in}+\frac{4\alpha_0 m^2}{\phi_0t^{-m+2}}
-\frac{4n^2m^2\phi_0^2}{t^{2(m+2)}}\left[{1\over 12\lambda^2}+{2\alpha_0 \over \phi_0^3t^{-3m}}\left(\frac{1}{\lambda^2}-\frac{\Lambda}{6\lambda^4}\right)\right]\\&\hspace{0.3 in}+\frac{4n^2m(m+3-2n)\phi_0}{t^{(m+4)}}\left[-{\phi_0t^{-m}\over 12\lambda^2}+{\alpha_0 \over \phi_0^2t^{-2m}}\left(\frac{1}{\lambda^2}-\frac{\Lambda}{6\lambda^4}\right)\right]\\&\hspace{0.3 in}
+12\gamma_0\left[\frac{32n^5(2n-1)(2n-2)(2n-3)}{t^8}+\frac{256n^5(2n-1)^2(2n-2)}{t^8}-\frac{576n^6(2n-1)(2n-2)}{t^8}\right]\\&\hspace{0.3 in}+12\gamma_0\left[\frac{192n^5(2n-1)^3}{t^8}
-\frac{2160n^6(2n-1)^2}{t^8}+\frac{5088n^7(2n-1)}{t^8}-\frac{3120n^8}{t^8}\right]\\&\hspace{0.3 in}=-{\omega\rho_{0}\over c^{3(1+\omega)} t^{3n(1+\omega)}}-{m^2\phi_0^2\over 2t^{2(m+1)}} +{{\lambda^2\phi_0^2t^{-2m}}\over 2}-C_0.\end{split}\ee

\be\label{ME}\begin{split}& {2 \alpha_0  \over \phi_0t^{-m}}\left({3n^2\over t^2}-\Lambda\right)+\frac{6\alpha_0nm}{t^2}-\frac{24m\mathrm{}n^3\phi_0}{t^{m+4}}\left[-{\phi_0t^{-m}\over 24\lambda^2}+{\alpha_0 \over \phi_0^2t^{-2m}}\left(\frac{1}{2\lambda^2}-\frac{\Lambda}{12\lambda^4}\right)\right]\\&\hspace{0.3 in}+18\gamma_0\left[\frac{64n^6(2n-1)(2n-2)}{t^8}+\frac{96n^6(2n-1)^2}{t^8}-\frac{576n^7(2n-1)}{t^8}+\frac{480n^8}{t^8}\right] \\&\hspace{0.3 in}={\rho_{0}\over  c^{3(1+\omega)} t^{3n(1+\omega)}}+{m^2\phi_0^2\over 2t^{2(m+1)}} +{{\lambda^2\phi_0^2t^{-2m}}\over 2}-C_0.\end{split}\ee

\be\label{Mphi}\begin{split}&{ m(m+1)\phi_0\over t^{m+2}}-{3nm\phi_0\over t^{m+2}}+\frac{6\alpha_0n(2n-1)}{\phi_0^2t^{2-2m}}-{{2\alpha_0\Lambda}\over \phi_0^2t^{-2m}}-\frac{24n^3(2n-1)}{t^4}\left[-{\phi_0t^{-m}\over 24\lambda^2}+{\alpha_0 \over \phi_0^2t^{-2m}}\left(\frac{1}{2\lambda^2}-\frac{\Lambda}{12\lambda^4}\right)\right]\\&\hspace{0.3 in}+\frac{24n^4}{t^4}\left[-{\phi_0t^{-m}\over 24\lambda^2}+{\alpha_0 \over \phi_0^2t^{-2m}}\left(\frac{1}{2\lambda^2}-\frac{\Lambda}{12\lambda^4}\right)\right]+\lambda^2\phi_0t^{-m}=0.\end{split}\ee
Now, equation \eqref{Mphi} may only be satisfied for $n={1\over 2}$, provided $m =0$, $\alpha_0={\phi_0^3\over 6}$ and $\Lambda=3\lambda^2$. Note that, $m = 0$ simply implies $\phi=\phi_0 ~\text{(constant)}$. This means a viable matter dominates era may be realized if the scalar field seizes to evolve any further. This as such does not create any problem, since it has been revealed that a possible transition from decelerated to accelerated pressureless dust dominated epoch, admitting phantom phases, naturally appear in the framework of $F(R,\mathcal G)$, without opting for a scalar field \cite{I82,I83}.

\subsection{Radiation dominated era:}

In the radiation dominated era, $\omega={1\over3}$, and hence $\rho = {\rho_{r0}\over a^4}$, in view of equation \eqref{bi}, where, $\rho_{r0}$ is the amount of radiation density available in the present universe. As mentioned, equation \eqref{Mphi} is satisfied only for $n={1\over 2}$ and $m = 0$, in which case, the first two equations \eqref{MZ} and\eqref{ME} reduce to,

\be \begin{split}\label{zrad}&
{\alpha_0\over 2\phi_0 t^2}+ {2\alpha_0 \Lambda\over \phi_0} + {585\gamma_0\over 4 t^8}={\rho_{r0}\over 3c^2 t^2} -{\lambda^2\phi_0^2\over 2}+C_0, \end{split}\ee
and
\be\begin{split}\label{Erad}&{3\alpha_0\over 2\phi_0 t^2}- {2 \alpha_0 \Lambda\over \phi_0} + {135\gamma_0\over 4 t^8} = {\rho_{r0}\over c^2 t^2} +{\lambda^2\phi_0^2\over 2}-C_0.\end{split}\ee
The above two equations are satisfied once we can neglect $t^{-8}$ term associated with $(\mathcal{G})^2$. However, since the universe enters hot big bang era at around ($t \sim 10^{-18\pm 6}$s), so in the early radiation era $t^{-8}$ term gives dominant contribution. Hence, the cosmic evolution of early radiation era will typically be different from the standard model, which might tell upon structure formation. Nonetheless, at the late stage, this term may be safely neglected. In that case, we are simply left with Friedmann equation along with an effective cosmological constant, and the above pair of equations are satisfied for $C_0 ={\lambda^2\phi_0^2\over 2} + {2\alpha_0\Lambda \over \phi_0}={3\lambda^2\phi_0^2\over 2}$, and $\rho_{r0}=\frac{3\alpha_0c^2}{2\phi_0}={\phi_0^2 c^2\over 4}$. Note that, one can add a term $b_0 \mathcal{G}^{n_2}$ at this stage with $n_2 < {1\over 2}$. Unfortunately, the field equations in no way are satisfied, unless $b_0 = 0$. We remind, it was demonstrated that the form $F(\mathcal{G}) = a_0 \mathcal{G}^{n_1} + b_0 \mathcal{G}^{n_2}$ can unify early inflationary era (if $n_1 > 1$) with late-time accelerated expansion, (if $n_2 < {1\over 2}$) \cite{I83}. Since a viable radiation era does not admit $n_2 < {1\over 2}$, late-time acceleration in the pressureless dust era with powers of Gauss-Bonnet term is questionable. Further, since we have already shown that no other than $a\propto \sqrt t$ satisfies the field equations, a viable ($a\propto t^{2\over 3}$, as in the standard model of cosmology) early pressureless dust era also remains obscure.

\section{\bf{Concluding remarks:}}

It was argued that $F(\mathcal{G}) = a_0 \mathcal{G}^{n_1} + b_0 \mathcal{G}^{n_2}$ can unify early inflationary phase (if $n_1 > 1$) and late-time accelerating phase, (if $n_2 < {1\over 2}$). However, in no way one can disregard a dilatonic coupled Gauss-Bonnet term in the very early epoch, since it is an artefact of different string inspired theories. Taking into account a scalar-tensor theory of gravity, associated with dilatonic coupled Gauss-Bonnet term and a Gauss-Bonnet squared term, we study inflation. Inflation was triggered at a very early epoch, when all the fields but gravity are quantized. Therefore, prior to study inflationary dynamics from classical field equation, it is suggestive to construct the quantum counterpart of the theory and to see if a legitimate semiclassical wave function emerges. This we have performed and studied inflation thereafter.\\

For a theory having additional degrees of freedom in the form of coupling parameters, classical field equations are usually reduced to Friedmann-like equations under the choice of additional hierarchy of flow parameters. Earlier, for a single additional degree of freedom, the technique worked nicely. However, in the present situation we had a couple of independent coupling parameters requiring two additional choices. This reduced the field equations substantially, without reflecting the contribution from Gauss-Bonnet-dilatonic term and the Gauss-Bonnet squared term. We have therefore followed a different method to reduce the extremely complicated set of field equations considerably, by choosing effective potentials. We have studied three independent cases for three different choices of the redefined potentials, and find that the inflationary parameters remain almost unaltered. The last case have been studied both in Jordan as well as Einstein's frames and no remarkable change in the inflationary parameters has been noticed. It has been shown that it is impossible to meet the present observational constraint on tensor to scalar ratio ($r < .032$), which might even be reduced in future experiments. It is noteworthy that our results are independent of the choice of the parameters of the theory.\\

 Further, it is found that a viable radiation era is possible only at the late stage, prior to the pressureless dust era, provided the scalar field ceases to evolve and Gauss-Bonnet squared term is neglected. This might tell upon structure formation and also on the formation of CMBR at right epoch. The matter dominated era also evolves as $a \propto \sqrt t$, instead of $a \propto t^{2\over 3}$. Further, since a viable radiation era does not admit term like $\mathcal{G}^{n_2}$ for $n_2 < {1\over 2}$, hence late stage accelerated expansion remains obscure. In a nut shell, present study reveals that the theory does not quite justify cosmological evolution in different regimes.\\

\noindent
\textbf{Data Availability Statement:} The authors declare that all data supporting the findings of this study are available within the article.

\noindent
\textbf{CRediT authorship contribution statement:}
Dalia Saha: Computation, Cross-checking, Writing - review and editing. Jyoti Prasad Saha: Computation, Cross-checking, Writing - review and editing. Abhik Kumar Sanyal: Cross-checking, Supervision, Writing - original draft, Writing - review and editing.\\

\noindent
\textbf{Declaration of competing interest}
The authors declare that they have no known competing financial interests or personal relationships that could have appeared to influence the work reported in this paper.\\

\noindent
\textbf{Funding:}
We also declare that no funding is available for the present research from any organization.\\

\appendix
\numberwithin{equation}{section}
\section{Appendix A: Phase space structure, Dirac's algorithm:}
In fact there are two known formalisms which can handle such singular action being associated with higher order terms, towards expressing it in canonical form. The first is the well known Dirac formalism. It works to produce a viable quantum theory, if applied after removing the total derivative terms present in the action, as we already have done, and expressed it in equation \eqref{A2} and then finally in \eqref{A3}, under a change of variable. The other technique, known as `modified Horowitz' formalism' (MHF), bypasses Dirac's constraint analysis, and works as well to produce identical phase-space structure. There is yet another formalism presented by Buchbinder and his collaborators, which we have recently explored (D.Saha and A.K. sanyal, Perusing Buchbinder-Lyakhovich Canonical Formalism for the Higher-Order Theory of Gravity, Universe, 9, 48, (2023)), and found viable to produce identical phase-space structure too. However, we are not considering it here. It is duly required to mention that all other techniques which do not take care of the divergent terms (inclusive of Dirac's formalism), also produce phase-space structure of the Hamiltonian, which although different, are canonically equivalent. However, such equivalence remains obscure in the quantum domain, due to non-linearity, and the associated quantum counterparts run into problem \cite{Aks}. Here, as mentioned we shall follow Dirac's constraint analysis starting from the action \eqref{A3}, and in the Appendix B explore MHF, to prove equivalence between the two. The point Lagrangian in connection with the action \eqref{A3} takes the following form,

\be\begin{split}\label{L} L& = -{6\alpha(\phi) N}\Big({x^2\over 4\sqrt z} - k\sqrt z + {\Lambda\over 3} z^{3\over 2}\Big)-{3\alpha'(\phi) \dot\phi x\sqrt z}-\frac{\beta'x\dot\phi}{\sqrt z}\bigg{(}\frac{x^2}{z}+12k \bigg{)} + N z^{3\over 2}\Big({1\over 2 N^2}\dot\phi^2 - V(\phi)\Big)\\&
+ 144\gamma(\phi)\Bigg\{{(x^2 + 4 k z)^2\dot x^2\over 16 N z^{9\over 2}}+{\gamma'(\phi)\dot\phi\over \gamma(\phi)}\Big({x^7\over 112 z^{11\over 2}} + {k x^5\over 10 z^{9\over 2}}+ {k^2 x^3\over 3 z^{7\over 2}}\Big)- N\Big({15 x^8 \over 448 z^{13\over 2}} + {13 k x^6\over 40 z^{11\over 2}} + {11 k^2 x^4\over 12 z^{9\over 2}}\Big)\Bigg\}\\& \hspace{0.4 in}+ u\left({\dot z\over N} - x\right).\end{split}\ee
Note that, as a result of the change of variable \eqref{zx}, expression $\big{(}{\dot z\over N} - x =0\big{)}$ turns out to be a constraint, which has been introduced in the above point Lagrangian through the Lagrange multiplier $u$. The canonical momenta therefore are,

\be\label{p1}\begin{split} & p_{\phi}=-{3\alpha'x\sqrt z}+{z^{3\over 2}\dot\phi\over N}+144\gamma'(\phi)\left({x^7\over 112 z^{11\over 2}} + {k x^5\over 10 z^{9\over 2}}+ {k^2 x^3\over 3 z^{7\over 2}}\right)-\frac{\beta'x}{\sqrt z}\bigg{(}\frac{x^2}{z}+12k \bigg{)},\\&p_x =\left[{288\gamma\over N}\left({x^4\over 16 z^{9\over 2}}+{k x^2\over 2 z^{7\over 2}}+{k^2\over z^{5\over 2}}\right) \right]\dot x,\hspace{0.4 in} p_z = {u\over N}, \hspace{0.4 in} p_N = 0 = p_u, \end{split}\ee
and the resulting constrained Hamiltonian reads as,

\be\label{Hp1}\begin{split} H_{c}= & {Np_x^2\over {576\gamma \left({x^4\over 16 z^{9\over 2}}+{k x^2\over 2 z^{7\over 2}}+{k^2\over z^{5\over 2}}\right)}}+\dot z{u\over N}+{Np_{\phi}^2\over 2z^{3\over 2}}\\&+\left[{3N\alpha' \over z}-{144N \gamma'\over z^{3\over 2}}\left({x^7\over 112 z^{11\over 2}} + {k x^5\over 10 z^{9\over 2}}+ {k^2 x^3\over 3 z^{7\over 2}}\right)\right]p_{\phi}+\bigg{(}\frac{x^2}{z}+12k \bigg{)}\left[\frac{N\beta'x p_{\phi}}{ z^2}+\frac{3N\alpha'\beta' x^2}{z^{3\over 2}}\right]\\&-\left({x^7\over 112 z^{11\over 2}} + {k x^5\over 10 z^{9\over 2}}+ {k^2 x^3\over 3 z^{7\over 2}}\right)\left[{432\alpha'\gamma' xN\over z}+{288N\beta'\gamma' x\over z}\bigg{(}\frac{x^2}{z}+12k \bigg{)}\right]\\&+{9\alpha'^2 x^2N\over 2\sqrt z}
+\frac{N\beta'^2x^2}{2 z^{5\over 2}}\bigg{(}\frac{x^2}{z}+12k \bigg{)}^2+{10368N\gamma'^2\over z^{3\over 2}}\left({x^7\over 112 z^{11\over 2}} + {k x^5\over 10 z^{9\over 2}}+ {k^2 x^3\over 3 z^{7\over 2}}\right)^2\\&+ {6\alpha N}\left({x^2\over 4\sqrt z} - k\sqrt z + {\Lambda\over 3} z^{3\over 2}\right) + 36\gamma Nx^4 \left({15x^4\over 112 z^{13\over 2}}+{13k x^2\over 10z^{11\over 2}}+{11k^2\over 3 z^{9\over 2}}\right)+ NVz^{3\over 2}-\dot z{u\over N}+u x.\end{split}\ee
The definitions of momenta \eqref{p1} reveal that there exists following pair of primary constraints viz,

\be\label{constraints}\phi_1 = Np_z-u \approx 0,~ \phi_2 = p_u \approx 0,\ee
which involve Lagrange multipliers or their conjugates. Note that the constraint $\phi_3 = p_N$ associated with lapse function $N$ is non-dynamical, and so vanishes strongly. Thus it has been disregarded. The above two primary constraints \eqref{constraints} are second class, since their Poisson bracket with other constraints do not vanish. In two possible manner these second-class constraints may be tackled. Firstly, using arbitrary Lagrange multipliers, the constraints are inserted into the Hamiltonian, which may be determined explicitly. This helps to solve consistency equations, since $det |{\phi_i,\phi_j}| \neq 0$. Second, introducing Dirac bracket, in which case the constraints are disregarded. Note that the correct commutation relations required to build the quantum theory follow from Dirac brackets, therefore let us compute Dirac brackets first, thereafter follow the first method, which is straight forward. In phase-space, the Dirac bracket of two functions $f_1$ and $f_2$ is defined as (we use $DB$, and $PB$ in the suffix to denote Dirac and Poisson brackets respectively).

\be \big\{f_1,f_2\big\}_{DB} = \big\{f_1,f_2\big\}_{PB} - \sum_{ij}\big\{f_1,\phi_i\big\}_{PB}M^{-1}_{ij}\big\{\phi_j,f_2\big\}_{PB},\ee
where  $M^{-1}_{ij}$ denotes the inverse of the matrix $M_{ij} = \big\{\phi_i,\phi_j\big\}_{PB}$. In the present situation, the matrix and its inverse are simply
\be M_{ij} = \left(
            \begin{array}{cc}
              0 & -1 \\
              1 & 0 \\

            \end{array}
          \right)
~~ \mathrm{and}~~ M^{-1}_{ij} = \left(
            \begin{array}{cc}
              0 & 1 \\
              -1 & 0 \\

            \end{array}
          \right)\ee
and hence the Dirac bracket takes the following form:

\be \big\{f_1,f_2\big\}_{DB} = \big\{f_1,f_2\big\}_{PB} + \sum_{ij}\epsilon_{ij}\big\{f_1,\phi_i\big\}_{PB}\big\{\phi_j,f_2\big\}_{PB},\ee
where $\epsilon_{ij}$ is the Levi-Civita symbol. One can now compute Dirac bracket in a straightforward manner to find,

\be\begin{split} \{z,p_z\}_{DB} &= \{z,p_z\}_{PB} + \epsilon_{11}\{z,\phi_1\}_{PB}\{\phi_1,p_z\}_{PB} + \epsilon_{12}\{z,\phi_1\}_{PB}\{\phi_2,p_z\}_{PB}\\&\hspace{2cm}+ \epsilon_{21}\{z,\phi_2\}_{PB}\{\phi_1,p_z\}_{PB} + \epsilon_{22}\{z,\phi_2\}_{PB}\{\phi_2,p_z\}_{PB}\\&
      = \{z,p_z\}_{PB} = 1,\end{split}\ee
since, $\{\phi_i, p_z\}_{PB} = 0$. In the same way it is possible to show that: $\{x,p_x\}_{DB} = \{x,p_x\}_{PB} = 1$,  $\{z,p_x\}_{DB} = \{z,p_x\}_{PB} = 0$,$\{p_z,p_x\}_{DB} = \{p_z,p_x\}_{PB} = 0$. As a result, the standard commutation relations viz. $[\hat{z}, \hat{p}_z] = i\hbar$, $[\hat{z}, \hat{p}_x] = 0$, hold for canonical quantization. The equality holds because $\phi_2$ vanishes strongly. So, in this prescription, it may be disregarded, as mentioned. However, this is not possible while following the first prescription, because in that case the Lagrange multipliers would remain undetermined. We now switch over to the standard procedure as already mentioned. So we substitute the two constraints, and as a result the modified primary Hamiltonian reads as,

\be H_{p1} = H_c + u_1(Np_z - u) + u_2p_u \ee
In the above, we introduced the Lagrange multipliers $u_1$, $u_2$, and find that the Poisson brackets $\{x, p_x\} = \{z, p_z\} = \{u, p_u\} = \{\phi, p_{\phi}\} = 1$, hold.
The primary Hamiltonian therefore takes the following form,
\be\label{Hpc}\begin{split}  H_{p1}= &N\Bigg{[}{p_x^2\over {576\gamma \left({x^4\over 16 z^{9\over 2}}+{k x^2\over 2 z^{7\over 2}}+{k^2\over z^{5\over 2}}\right)}}+{p_{\phi}^2\over 2z^{3\over 2}}+\left[{3\alpha' x \over z}-{144 \gamma'\over z^{3\over 2}}\left({x^7\over 112 z^{11\over 2}} + {k x^5\over 10 z^{9\over 2}}+ {k^2 x^3\over 3 z^{7\over 2}}\right)\right]p_{\phi}\\&+\bigg{(}\frac{x^2}{z}+12k \bigg{)}\left[\frac{\beta'x p_{\phi}}{ z^2}+\frac{3\alpha'\beta' x^2}{z^{3\over 2}}\right]-\Big({x^7\over 112 z^{11\over 2}} + {k x^5\over 10 z^{9\over 2}}+ {k^2 x^3\over 3 z^{7\over 2}}\Big)\left[{432\alpha'\gamma' x\over z}+{288\beta'\gamma' x\over z}\bigg{(}\frac{x^2}{z}+12k \bigg{)}\right]\\&+{9\alpha'^2 x^2N\over 2\sqrt z}+\frac{\beta'^2x^2}{2 z^{5\over 2}}\bigg{(}\frac{x^2}{z}+12k \bigg{)}^2+{10368\gamma'^2\over z^{3\over 2}}\left({x^7\over 112 z^{11\over 2}} + {k x^5\over 10 z^{9\over 2}}+ {k^2 x^3\over 3 z^{7\over 2}}\right)^2\\&+ {6\alpha}\left({x^2\over 4\sqrt z} - k\sqrt z + {\Lambda\over 3} z^{3\over 2}\right)+36\gamma x^4 \left({15x^4\over 112 z^{13\over 2}}+{13k x^2\over 10z^{11\over 2}}+{11k^2\over 3 z^{9\over 2}}\right)+Vz^{3\over 2}\Bigg{]}\\&+u_1\big{(}Np_z-u\big{)}+u_2 p_u+ux.\end{split}\ee
Now since constraint should not change with time, in the sense of Dirac, therefore,

\be\begin{split}&\label{phi12} \dot\phi_1=\{\phi_1,H_{p1}\}=-u_2-N{\partial H_{p1}\over \partial z}\approx 0\Rightarrow u_2=-N{\partial H_{p1}\over \partial z};
\\& \dot\phi_2=\{\phi_2,H_{p1}\}=-x+u_1\approx 0\Rightarrow u_1=x. \end{split}\ee
Thus, both $u_1$ and $u_2$ are determined, and that too in such a manner, that the last term ($ux$) appearing in \eqref{Hpc} gets cancelled. The primary Hamiltonian therefore becomes:

\be\label{Hp1c}\begin{split}  H_{p2} = &N\Bigg{[}x p_z+{p_x^2\over{ 576\gamma \left({x^4\over 16 z^{9\over 2}}+{k x^2\over 2 z^{7\over 2}}+{k^2\over z^{5\over 2}}\right)}}+{p_{\phi}^2\over 2z^{3\over 2}}+\left[{3\alpha' x\over z}-{144 \gamma'\over z^{3\over 2}}\left({x^7\over 112 z^{11\over 2}} + {k x^5\over 10 z^{9\over 2}}+ {k^2 x^3\over 3 z^{7\over 2}}\right)\right]p_{\phi}\\&+\bigg{(}\frac{x^2}{z}+12k \bigg{)}\left[\frac{\beta'x p_{\phi}}{ z^2}+\frac{3\alpha'\beta' x^2}{z^{3\over 2}}\right]-\Big({x^7\over 112 z^{11\over 2}} + {k x^5\over 10 z^{9\over 2}}+ {k^2 x^3\over 3 z^{7\over 2}}\Big)\left[{432\alpha'\gamma' x\over z}+{288\beta'\gamma' x\over z}\bigg{(}\frac{x^2}{z}+12k \bigg{)}\right]\\&+{9\alpha'^2 x^2\over 2\sqrt z}+\frac{\beta'^2x^2}{2 z^{5\over 2}}\bigg{(}\frac{x^2}{z}+12k \bigg{)}^2+{10368\gamma'^2\over z^{3\over 2}}\left({x^7\over 112 z^{11\over 2}} + {k x^5\over 10 z^{9\over 2}}+ {k^2 x^3\over 3 z^{7\over 2}}\right)^2\\&+ {6\alpha}\left({x^2\over 4\sqrt z} - k\sqrt z + {\Lambda\over 3} z^{3\over 2}\right)
+36\gamma x^4 \left({15x^4\over 112 z^{13\over 2}}+{13k x^2\over 10z^{11\over 2}}+{11k^2\over 3 z^{9\over 2}}\right)+Vz^{3\over 2}\Bigg{]}-Np_u{\partial H_{p1}\over \partial z}.\end{split}\ee
Again following the prescription that the constraint should not change with time in the sense of Dirac, we find

\be\label{Dphi2}\dot\phi_2=\{\phi_2,H_{p2}\}=N{\partial H_{p1}\over \partial z}-N\bigg{[}{\partial H_{p1}\over \partial z}-Np_u{\partial^2 H_{p1}\over \partial z^2}\bigg{]}\approx 0\Rightarrow p_u=0. \ee
The system is now closed, the Hamiltonian no longer involve constraints, and finally reads as,

\be\label{Hp1c}\begin{split} & H=N\Bigg{[}x p_z+{p_x^2\over{ 576\gamma \left({x^4\over 16 z^{9\over 2}}+{k x^2\over 2 z^{7\over 2}}+{k^2\over z^{5\over 2}}\right)}}+{p_{\phi}^2\over 2z^{3\over 2}}+\left[{3\alpha' x\over z}-{144 \gamma'\over z^{3\over 2}}\left({x^7\over 112 z^{11\over 2}} + {k x^5\over 10 z^{9\over 2}}+ {k^2 x^3\over 3 z^{7\over 2}}\right)\right] p_{\phi}\\&+\bigg{(}\frac{x^2}{z}+12k \bigg{)}\left[\frac{\beta'x p_{\phi}}{ z^2}+\frac{3\alpha'\beta' x^2}{z^{3\over 2}}\right]-\left({x^7\over 112 z^{11\over 2}} + {k x^5\over 10 z^{9\over 2}}+ {k^2 x^3\over 3 z^{7\over 2}}\right)\left[{432\alpha'\gamma' x\over z}+{288\beta'\gamma' x\over z}\bigg{(}\frac{x^2}{z}+12k \bigg{)}\right]\\&+{9\alpha'^2 x^2\over 2\sqrt z}
+\frac{\beta'^2x^2}{2 z^{5\over 2}}\bigg{(}\frac{x^2}{z}+12k \bigg{)}^2+{10368\gamma'^2\over z^{3\over 2}}\left({x^7\over 112 z^{11\over 2}} + {k x^5\over 10 z^{9\over 2}}+ {k^2 x^3\over 3 z^{7\over 2}}\right)^2\\&+ {6\alpha}\left({x^2\over 4\sqrt z} - k\sqrt z + {\Lambda\over 3} z^{3\over 2}\right)
+36\gamma x^4 \left({15x^4\over 112 z^{13\over 2}}+{13k x^2\over 10z^{11\over 2}}+{11k^2\over 3 z^{9\over 2}}\right)+Vz^{3\over 2}\Bigg{]}=N\mathcal{H}.\end{split}\ee
It is important to note that the diffeomorphic invariance $H = N\mathcal{H}$ is also established in the process. The action (\ref{A3}) may now be expressed in canonical ADM form ($k = 0$) as,
\begin{center}
    \be \label{ADMH} A=\int\bigg{(}\dot z p_z+\dot x p_x+\dot\phi p_{\phi}-N\mathcal{H}\bigg{)}dt d^3x=\int\bigg{(}\dot{h_{ij}}\pi^{ij}+\dot{K_{ij}}{\Pi}^{ij}+\dot\phi p_{\phi}-N\mathcal{H}\bigg{)}dt d^3x,\ee
\end{center}
where momenta $\pi^{ij}$ and $\Pi^{ij}$ are canonically conjugate to $h_{ij}$ and $K_{ij}$ respectively. Before we close this subsection, let us proceed to find explicit expression of the momentum $p_z$, which was originally appeared in the action \eqref{Hp1} as a constraint, from the Hamilton's equation. It will be required for later consideration. To avoid complications, let us express the Hamiltonian \eqref{Hp1c} for $k=0$ as,

\be\label{Hp2c}\begin{split} \mathcal{H} = &\Bigg{[}x p_z+{p_x^2\over{36\Big({\gamma x^4\over z^{9\over 2}}\Big)}}+{p_{\phi}^2\over 2z^{3\over 2}}+{3\alpha' x p_{\phi}\over z}+{\beta' x^3 p_{\phi}\over z^3}-{144\gamma'x^7 p_{\phi}\over 112 z^{7}}+{3\alpha'\beta'x^4 \over z^{5\over 2}}-{432\alpha'\gamma'x^8\over 112 z^{13\over 2}}-{288\beta'\gamma'x^{10} \over 112 z^{15\over 2}}\\&+{9\alpha'^2 x^2\over 2\sqrt z}+\frac{\beta'^2x^6}{2 z^{9\over 2}}+{10368\gamma'^2\over z^{3\over 2}}\Big({x^7\over 112 z^{11\over 2}}\Big)^2+ {6\alpha}\left({x^2\over 4\sqrt z} + {\Lambda\over 3} z^{3\over 2}\right)+36\gamma x^4 \left({15x^4\over 112 z^{13\over 2}}\right)+Vz^{3\over 2}\Bigg{]}.\end{split}\ee
Now Hamilton's equations are

\be\label{px1}\begin{split}&\dot x=\frac{p_x}{18\Big({\gamma x^4 \over {z}^{9\over2}}\Big)},~~~~\dot z=x,~~~~\dot\phi=\frac{p_{\phi}}{z^{3\over 2}}+\frac{3\alpha'x}{z}+\frac{\beta'x^3}{z^3}-144\frac{\gamma' x^7}{112z^7},\hspace{0.7 in}\dot p_{\phi}=0,\\&
\dot p_x=\bigg{(}-p_z+\frac{ p_x^2{z}^{9\over2}}{9\gamma x^5}-\frac{3\alpha'p_\phi}{z}-\frac{3\beta'x^2p_\phi}{z^3}+\frac{9\gamma'x^6 p_\phi}{z^7}-{12\alpha'\beta'x^3 \over z^{5\over 2}}+\frac{216\alpha'\gamma'x^7}{7z^{13\over2}}+{2880\beta'\gamma'x^{9} \over 112 z^{15\over 2}}\\&\hspace{1.1 in}-\frac{9\alpha'^2x}{\sqrt z}-\frac{3\beta'^2x^5}{ z^{9\over 2}}-\frac{81{\gamma'^2}x^{13}}{7z^{25\over2}}-\frac{3\alpha x}{\sqrt z}-\frac{270\gamma x^7}{7z^{13\over2}}\bigg{)},
\\& \dot p_z=\Bigg(-\frac{9 p_x^2{z^{{7\over 2}}}}{72{\gamma x^4}}+\frac{3p_{\phi}^2}{4z^{5\over 2}}+\frac{3 p_{\phi}\alpha'x }{z^2}+\frac{3p_\phi\beta'x^3}{z^4}-\frac{9p_{\phi}\gamma'x^7}{z^8}+\frac{15\alpha'\beta' x^4}{2z^{7\over 2}}-\frac{135\beta'\gamma' x^{10}}{7z^{17\over 2}}-\frac{351\alpha'\gamma' x^8}{28z^{15\over 2}}\\& \hspace{0.5 in}
+\frac{9\alpha'^2 x^2}{4z^{3\over 2}}+\frac{9\beta'^2 x^6}{4z^{11\over 2}}+\frac{2025\gamma'^2 x^{14}}{196z^{27\over 2}}+\frac{3\alpha x^2}{4z^{3\over 2}}-3\alpha\Lambda z^{1\over 2}+\frac{1755\gamma x^8}{56z^{15\over 2}}-\frac{3V z^{1\over 2}}{2}\Bigg).\end{split}\ee
From the first expression of the above set of equations, we find,

\be\label{px2} p_x=18\left(\frac{\gamma\dot z^4\ddot z}{z^{9\over2}}\right);~~ \mathrm{and~hence}~~\dot p_x=\frac{18\gamma'\dot\phi\dot z^4\ddot z}{z^{9\over2}}+\frac{18\gamma\dot z^4\dddot z}{z^{9\over2}}+\frac{72\gamma\dot z^3\ddot z^2}{z^{9\over2}}-\frac{81\gamma\dot z^5\ddot z}{z^{11\over2}}.\ee
Equating $\dot p_x$ obtained in equations \eqref{px1} and \eqref{px2}, we can express $p_z$ as,
\be\label{pz}\begin{split}& p_z = -18\gamma\bigg{(}\frac{2\dot z^3\ddot z^2}{z^{9\over2}}+\frac{15\dot z^7}{7z^{13\over2}}+\frac{\dot z^4\dddot z^2}{z^{9\over2}}-\frac{9\dot z^5\ddot z}{2z^{11\over2}}\bigg{)}+18\gamma'\dot\phi\bigg{(}\frac{\dot z^6}{2z^{11\over2}}-\frac{\ddot z\dot z^4}{z^{9\over2}}\bigg{)}-3\alpha'\dot\phi\sqrt z-\frac{3\alpha\dot z}{\sqrt z}\\& \hspace{1.1 in}-\frac{3\beta'\dot\phi\dot z^2}{z^{3\over 2}}-\frac{90\beta'\gamma'\dot z^9}{7z^{17\over 2}}+\frac{180\beta'\gamma'\dot z^9}{7z^{15\over 2}}.\end{split}\ee
Thus, \eqref{p1} and \eqref{pz} constitute the expressions for the whole set of momenta ($p_x$, $p_z,$ and $p_\phi$).

\subsection{Canonical quantization:}

Since the Hamiltonian (\ref{Hp1c}) has been constructed, canonical quantization is possible in a straight forward manner,

\be \label{qh} \begin{split} {i\hbar z^{-{9\over 2}}}\frac{\partial\Psi}{\partial z} = &-\frac{\hbar^2}{36x[\gamma{(}x^2+4kz{)}^2]} \bigg{(}\frac{\partial^2}{\partial x^2} +\frac{n}{x}\frac{\partial}{\partial x}\bigg{)}\Psi-\frac{\hbar^2}{2xz^6}\frac{\partial^2\Psi}{\partial \phi^2}+{3 \widehat{\alpha'p_{\phi}}\over z^{11\over 2}}\Psi+\frac{1}{ z^{13\over 2}}\bigg{(}\frac{x^2}{z}+12k \bigg{)}\widehat{\beta' p_{\phi}}\psi\\&-{144}\left({x^6\over 112 z^{23\over 2}} + {k x^4\over 10 z^{21\over 2}}+ {k^2 x^2\over 3 z^{19\over 2}}\right)\widehat{\gamma'p_{\phi}}\Psi +\Bigg{[}\frac{3x\alpha'\beta'}{z^6}\Big({ x^2\over z }+12k\Big)\\&-\left({x^7\over 112 z^{11}} + {k x^5\over 10 z^{10}}+ {k^2 x^3\over 3 z^{9}}\right)\left\{{432\alpha'\gamma'}+{288\beta'\gamma'}\Big({ x^2\over z }+12k\Big)\right\}+{9\alpha'^2 x\over 2z^5}\\& +{\beta'^2 x\over 2z^7}\left({x^2\over z} +12k\right)^2+ {10368\gamma'^2\over xz^{6}}\left({x^7\over 112 z^{11\over 2}} + {k x^5\over 10 z^{9\over 2}}+ {k^2 x^3\over 3 z^{7\over 2}}\right)^2+6\alpha\left({x\over {4z^5}}-{k\over{ xz^4}}+{\Lambda\over 3xz^3}\right) \\&+ 36\gamma x^3 \left({15x^4\over 112 z^{11}}+{13k x^2\over 10z^{10}}+{11k^2\over 3 z^9}\right)+{V\over xz^{3}}\Bigg{]}\Psi , \end{split}\ee
where, in order to resolves some operator ordering ambiguities, the operator ordering index $n$ is introduced. The above quantum expression is still devoid of some additional operator ordering ambiguities, particularly between ($\alpha'$ and $p_\phi$), ($\beta'$ and $p_\phi$) as well as between ($\gamma'$ and $p_\phi$). These ambiguities may be resolved only after specifying the forms of $\alpha(\phi),~ \beta(\phi)$ and $\gamma(\phi)$. The specific forms or relations amongst the coupling parameters $\alpha(\phi)$, $\beta(\phi)$, $\gamma(\phi)$ and $V(\phi)$ may be obtained once slow-roll inflation is invoked, so that on large scale, successful generation of almost scale-invariant perturbation from quantum fluctuations of $\phi$ is achieved. However, the classical de-Sitter solutions \eqref{aphi} already determined the forms of these coupling parameters, which we use to remove rest of the operator ordering ambiguities, and furnish the quantum counterpart of the present theory as,

\be \label{qh1} \begin{split} {i\hbar z^{-{9\over 2}}}\frac{\partial\Psi}{\partial z} = & -\frac{\hbar^2}{36\gamma x^5}\bigg{(}\frac{\partial^2}{\partial x^2} +\frac{n}{x}\frac{\partial}{\partial x}\bigg{)}\Psi-\frac{\hbar^2}{2xz^6}\frac{\partial^2\Psi}{\partial \phi^2}+\frac{3i\hbar \alpha_0}{z^{\frac{11}{2}}}\bigg{(}\frac{1}{\phi^2}\frac{\partial\Psi}{\partial \phi}-\frac{\Psi}{\phi^3}\bigg{)}-\frac{i\hbar x^2 \alpha_0\beta_0}{z^{\frac{15}{2}}}\bigg{(}\frac{1}{\phi^2}\frac{\partial\Psi}{\partial \phi}-\frac{\Psi}{\phi^3}\bigg{)}\\&+ \frac{i\hbar x^2 \beta_1}{z^{\frac{15}{2}}}\bigg{(}2\phi\frac{\partial\Psi}{\partial \phi}+{\Psi}\bigg{)} +\frac{9x}{2z^5} \bigg{(} \frac{\alpha_0^2}{\phi^4}\bigg{)}\Psi + \frac{ x^5}{2z^9}\left( {\alpha_0\beta_0\over \phi^2}-2\beta_1\phi\right)^2 \Psi-\frac{3x^3\alpha_0}{z^7\phi^2}\left({\alpha_0\beta_0\over \phi^2}-2\beta_1\phi\right)\Psi \\& +\bigg{[}\frac{3 x\alpha_0}{2z^5\phi}+\frac{135\gamma_0 x^7}{28 z^{11}}+\frac{2\Lambda\alpha_0}{ xz^3\phi} +\frac{1}{xz^3}\Big{(}{1\over 2}\lambda^2\phi^2-576\gamma_0\lambda^8\Big{)}\bigg{]}\Psi. \end{split}\ee
In the above, we have carefully performed Weyl symmetric ordering between $\{\alpha'$, $p_{\phi}\}$~ and $\{\beta'$, $p_{\phi}\}$, and since $\gamma(\phi) = \gamma_0$ is a constant, $\gamma'$ vanishes. Now, changing the variable ($z = \sigma^{2\over 11}$), the above modified Wheeler-de-Witt equation, looks like Schr\"odinger equation, viz.,

\be \label{qh2} \begin{split} {i\hbar}\frac{\partial\Psi}{\partial \sigma}&=-\frac{\hbar^2}{198\gamma x^5}\bigg{(}\frac{\partial^2}{\partial x^2} +\frac{n}{x}\frac{\partial}{\partial x}\bigg{)}\Psi-\frac{\hbar^2}{11x\sigma^{\frac{12}{11}}}\frac{\partial^2\Psi}{\partial \phi^2}+\frac{6i\hbar\alpha_0} {11\sigma}\bigg{(}\frac{1}{\phi^2}\frac{\partial\Psi}{\partial \phi}-\frac{\Psi}{\phi^3}\bigg{)}\\&\hspace{0.5 in}-\frac{2i\hbar x^2\alpha_0\beta_0}{11\sigma^{15\over 11}}\bigg{(}\frac{1}{\phi^2}\frac{\partial\Psi}{\partial \phi}-\frac{\Psi}{\phi^3}\bigg{)}+\frac{2i\hbar x^2\beta_1}{11\sigma^{15\over 11}}\bigg{(}2\phi\frac{\partial\Psi}{\partial \phi}+\Psi\bigg{)}+V_{e}\Psi=\widehat H_e \Psi, \end{split}\ee
where, $\sigma=z^{\frac{11}{2}}=a^{11}$ acts as `internal time parameter', and $\widehat H_e$ stands for the effective Hamiltonian operator. Clearly the very important role of the linear term in momentum ($xp_z$) that appeared in the Hamiltonian \eqref{Hp1c} has been exhibited. The effective potential $V_e$ in the above equation, is given by,\\
\be\begin{split} V_{e} = &\frac{9x}{11\sigma^{\frac{10}{11}}}\bigg{(}\frac{\alpha_0^2}{\phi^4}\bigg{)}  +\frac{x^{5}}{11\sigma^{\frac{18}{11}}}\left( {\alpha_0\beta_0\over \phi^2}-2\beta_1\phi\right)^2-\frac{6x^3 \alpha_0}{11\phi^2\sigma^{14\over 11}}\left( {\alpha_0\beta_0\over \phi^2}-2\beta_1\phi\right)\\&
\hspace{0.8 in}+\bigg{[}\frac{3\alpha_0x}{11\phi\sigma^{\frac{10}{11}}}+\frac{135\gamma_0 x^7}{154\sigma^2}+\frac{4\alpha_0\Lambda}{11x\phi\sigma^{6\over 11}}
+\frac{2}{11x\sigma^{\frac{6}{11}}}\Big{(}\frac{\lambda^2\phi^2}{2}-576\gamma_0\lambda^8\Big{)}\bigg{]}.\end{split} \ee

\subsection{Probabilistic interpretation:}

Having obtained quite a complicated quantized version of the theory under consideration, it is required to check hermiticity of the effective Hamiltonian operator $\widehat{H_e}$ presented in \eqref{qh2}. However, since the computation is straight forward and in a sense, `run of the machine', we place it in appendix C, where it has been proved that indeed $\widehat{H_e}$ is hermitian, provided the operator ordering index $n$ is set to $n = -5$. Thus, we proceed to write the continuity equation,
\begin{center}
\be \label{cont1}\frac{\partial\rho}{\partial\sigma}+\nabla. \textbf{J}=0, \ee
\end{center}
where, $\rho=\Psi^*\Psi$, and $\mathbf{J} = (J_x, J_z, J_\phi)$ are the probability density and current density respectively. In order to express the quantum equation in the above form \eqref{cont1}, it is required to find $\frac{\partial\rho}{\partial\sigma}$. A little algebra leads to the following equation:
\be\begin{split} \frac{\partial\rho}{\partial\sigma}= & -\frac{\partial}{\partial x}\Bigg[\frac{i\hbar}{198\gamma x^5 }\big{(}\Psi\Psi^*_{,x}-\Psi^*\Psi_{,x} \big{)}\Bigg]-\frac{i\hbar \big{(}\Psi\Psi^*_{,x}-\Psi^*\Psi_{,x} \big{)}  }{198\gamma x^6}\left(n+5\right)\\& -\frac{\partial}{\partial\phi}\bigg{[}\frac{i\hbar}{11x\sigma^{\frac{12}{11}}} \big{(}\Psi \Psi^*_{,\phi}-\Psi^*\Psi_{,\phi} \big{)}-\frac{6}{11\sigma}\Big{(}\frac{\alpha_0}{\phi^2}
-\frac{{\alpha_0\beta_0}x^2}{3\phi^2\sigma^{\frac{4}{11}}}+\frac{{2\beta_1}x^2\phi}{3\sigma^{\frac{4}{11}}}\Big{)}\Psi^*\Psi\bigg{]}, \end{split} \ee
which clearly reveals the fact that the continuity equation can only be written in its standard form,
\begin{center}
\be \frac{\partial\rho}{\partial\sigma}+\frac{\partial J_x}{\partial x}+\frac{\partial J_z}{\partial z}+\frac{\partial J_\phi}{\partial \phi}=0, \ee
\end{center}
again under the choice $n=-5$, where,
\be\begin{split}&
J_x = \frac{i\hbar}{198\gamma x^5}\big{(}\Psi\Psi^*_{,x}-\Psi^*\Psi_{,x} \big{)},\hspace{0.5 in} J_z = 0, \\&
J_{\phi} =  \frac{i\hbar}{11x\sigma^{\frac{12}{11}}} \big{(}\Psi \Psi^*_{,\phi}-\Psi^*\Psi_{,\phi} \big{)}-\frac{6}{11\sigma}\Big{(} \frac{\alpha_0}{\phi^2}-\frac{{\alpha_0\beta_0}x^2}{3\phi^2\sigma^{\frac{4}{11}}}+\frac{{2\beta_1}x^2\phi}{3\sigma^{\frac{4}{11}}}\Big{)}\Psi^*\Psi.
\end{split}\ee
We conclude at this end that the operator ordering index is fixed to $n = -5$ from the physical argument, that the effective Hamiltonian operator is hermitian and standard quantum mechanical probability interpretation holds. We remind that the variable $\sigma$ or the scale factor $a$ in disguise, plays the role of internal time parameter.

\subsection{Semiclassical approximation}

Since the quantum theory (\ref{qh2}) is found to be well behaved, it is necessary to check if such a theory results in the universe we live in. This may only be justified if the quantum equation transits over to the classical domain following a pertinent semiclassical approximation. In this connection the Hartle criterion for the selection of classical trajectories \cite{Hartle} is noteworthy: ``if the approximate wavefunction obtained following some appropriate semiclassical approximation is strongly peaked, then there exists correlations among the geometrical and matter degrees of freedom, and the emergence of classical trajectories (i.e. the classical universe) is expected, on the contrary, if it is not peaked, correlations are lost". In this subsection, we therefore proceed to find the semiclassical wavefunction. For convenience, we consider the equation (\ref{qh1}), instead of (\ref{qh2}), and express it as:

\begin{center}
\be \label{semi} \begin{split} &\frac{-\hbar^2 z^{\frac{9}{2}}}{36\gamma x^5}\bigg{(}\frac{\partial^2}{\partial x^2} +\frac{n}{x}\frac{\partial}{\partial x}\bigg{)}\Psi-\frac{\hbar^2}{2xz^{\frac{3}{2}}}\frac{\partial^2\Psi}{\partial \phi^2}-{i\hbar}\frac{\partial\Psi}{\partial z}+\frac{i\hbar} {z}\bigg{[}\frac{3\alpha_0}{\phi^2}-\frac{x^2\alpha_0\beta_0}{z^2\phi^2}+\frac{2x^2\beta_1\phi}{z^2}\bigg{]}\frac{\partial\Psi}{\partial \phi}\\& \hspace{1.47 in}-\frac{i\hbar}{z} \bigg{[}\frac{3\alpha_0}{\phi^3}-\frac{x^2\alpha_0\beta_0}{z^2\phi^3}-\frac{x^2\beta_1}{z^2}\bigg{]}\psi +\mathrm{V}(x,z,\phi)\Psi=0, \end{split}\ee
\end{center}
where
\be\begin{split}& \mathrm{V}(x,z,\phi)=\bigg{[}\frac{3 x}{2\sqrt z}\frac{\alpha_0}{\phi} +\frac{9x}{2\sqrt z} \frac{\alpha_0^2}{\phi^4}+ \frac{135\gamma_0 x^7}{28 z^{13\over 2}} +\frac{x^{5}}{2z^{\frac{9}{2}}}\left( {\alpha_0\beta_0\over \phi^2}-2\beta_1\phi\right)^2\\&\hspace{1.47 in}-\frac{3x^3\alpha_0}{\phi^2 z^{5\over 2}}\left( {\alpha_0\beta_0\over \phi^2}-2\beta_1\phi\right)+\frac{2\Lambda \alpha_0z^{\frac{3}{2}}}{x\phi}
+\frac{z^{\frac{3}{2}}}{x}\Big{(}{\lambda^2\phi^2\over 2}-576\gamma_0\lambda^8\Big{)}\bigg{]}\end{split} \ee
Equation (\ref{semi}) may be treated as time independent Schr{\"o}dinger equation with three variables ($x$, $z$, $\phi$). Let us therefore, seek the solution of equation (\ref{semi}) as,

\be\label{Psi}\Psi = \psi_0(x,z,\phi) e^{\frac{i}{\hbar}S(x,z,\phi)},\ee
where, we treat $\psi_0$ as a slowly varying function with respect to the phase $S$. We now expand $S$ in the power series of $\hbar$ as,
\be\label{S} S = S_0(x,z,\phi) + \hbar S_1(x,z,\phi) + \hbar^2S_2(x,z,\phi) + .... ,\ee
and thereafter compute,

\be\label{derivative}\begin{split} &\Psi_{,x} = \psi_{0,x}e^{\frac{i}{\hbar}S} + \frac{i}{\hbar}\Big[S_{0,x} + \hbar S_{1,x} + \hbar^2S_{2,x} + \mathcal{O}(\hbar)\Big]\psi_0e^{\frac{i}{\hbar}S};\\
&\Psi_{,xx} = \Psi_{0,xx}e^{\frac{i}{\hbar}S} + 2\frac{i}{\hbar}\Big[S_{0,x} + \hbar S_{1,x} + \hbar^2S_{2,x} + \mathcal{O}(\hbar)\Big]\psi_{0,x}e^{\frac{i}{\hbar}S} + \frac{i}{\hbar}\Big[S_{0,xx} + \hbar S_{1,xx} + \hbar^2S_{2,xx} + \mathcal{O}(\hbar)\Big]\psi_0e^{\frac{i}{\hbar}S}\\
&\hspace{1.2cm} - \frac{1}{\hbar^2}\Big[S_{0,x}^2 + \hbar^2 S_{1,x}^2 + \hbar^4S_{2,x}^4 + 2\hbar S_{0,x}S_{1,x} + 2\hbar^2 S_{0,x}S_{2,x} + 2\hbar^3 S_{1,x}S_{2,x} + \mathcal{O}(\hbar)\Big]\psi_0e^{\frac{i}{\hbar}S};\\
&\Psi_{,\phi} = \psi_{0,\phi}e^{\frac{i}{\hbar}S} + \frac{i}{\hbar}\Big[S_{0,\phi} + \hbar S_{1,\phi} + \hbar^2S_{2,\phi} + \mathcal{O}(\hbar)\Big]\psi_0e^{\frac{i}{\hbar}S};\\
&\Psi_{,\phi\phi} = \Psi_{0,\phi\phi}e^{\frac{i}{\hbar}S} + 2\frac{i}{\hbar}\Big[S_{0,\phi} + \hbar S_{1,\phi} + \hbar^2S_{2,\phi} + \mathcal{O}(\hbar)\Big]\psi_{0,\phi}e^{\frac{i}{\hbar}S}\\&\hspace{1.2cm} + \frac{i}{\hbar}\Big[S_{0,\phi\phi} + \hbar S_{1,\phi\phi} + \hbar^2S_{2,\phi\phi} + \mathcal{O}(\hbar)\Big]\psi_{0}e^{\frac{i}{\hbar}S}\\
&\hspace{1.0cm} - \frac{1}{\hbar^2}\Big[S_{0,\phi}^2 + \hbar^2 S_{1,\phi}^2 + \hbar^4S_{2,\phi}^4 + 2\hbar S_{0,\phi}S_{1,\phi} + 2\hbar^2 S_{0,\phi}S_{2,\phi} + 2\hbar^3 S_{1,\phi}S_{2,\phi} + \mathcal{O}(\hbar)\Big]\psi_{0}e^{\frac{i}{\hbar}S};\\
&\Psi_{,z} = \psi_{0,z}e^{\frac{i}{\hbar}S} + \frac{i}{\hbar}\Big[S_{0,z} + \hbar S_{1,z} + \hbar^2S_{2,z} + \mathcal{O}(\hbar)\Big]\psi_{0}e^{\frac{i}{\hbar}S}.\\\end{split}\ee
In the above, we use `comma' in the suffix to denote derivative. Now substituting all the above expressions (\ref{derivative}) in equation (\ref{semi}), and equating the coefficients of different powers of $\hbar$ to zero, we obtain the following set of equations (up to second order) as:
\be\begin{split}
&\frac{z^\frac{9}{2}}{36\gamma x^5}S_{0,x}^2 + \frac{S_{0,\phi}^2}{2xz^{\frac{3}{2}}} + S_{0,z}-\frac{1} {z}\bigg{[}\frac{3\alpha_0}{\phi^2}-\frac{x^2\alpha_0\beta_0}{z^2\phi^2}+\frac{2x^2\beta_1\phi}{z^2}\bigg{]} S_{0,\phi} + \mathrm{V}(x,z,\phi) = 0, \label{semi1}\end{split}\ee
\be\begin{split}
&\frac{z^\frac{9}{2}}{36\gamma x^5}\Big[\Big(i S_{0,xx} - 2S_{0,x}S_{1,x} + \frac{i}{x}n S_{0,x}\Big)\psi_0 + 2iS_{0,x}\psi_{0,x} \Big]
+\frac{1}{2xz^{3\over2}}\Big[\Big(i S_{0,\phi\phi} - 2S_{0,\phi}S_{1,\phi}\Big)\psi_0 + 2iS_{0,\phi}\psi_{0,\phi}\Big] \\& \hspace{0.4 in}-  S_{1,z}\psi_0
-\frac{i}{z}\left(\frac{3\alpha_0}{\phi^2}-\frac{x^2\alpha_0\beta_0}{z^2\phi^2}+\frac{2x^2\beta_1\phi}{z^2}\right)\Big[i S_{1,\phi}\psi_0+\psi_{0,\phi}\Big]-\frac{i}{z}\left(\frac{3\alpha_0}{\phi^3}
-\frac{x^2\alpha_0\beta_0}{z^2\phi^3}-\frac{x^2\beta_1}{z^2}\right ) = 0,\label{semi2}\end{split}\ee
\be\begin {split}
&\frac{z^\frac{9}{2}}{36\gamma x^5}\Big[\Big(i S_{1,xx} - S_{1,x}^2 - 2S_{0,x}S_{2,x} + \frac{i}{x}n S_{1,x}\Big)\psi_0 + \psi_{0,xx} + 2iS_{1,x}\psi_{0,x} + \frac{n}{x}\psi_{0,x}\Big] \\& \hspace{0.4 in}+\frac{1}{2xz^{3\over2}}\Big[\Big(i S_{1,\phi\phi} - S_{1,\phi}^2 - 2S_{0,\phi}S_{2,\phi}\Big)\psi_0 + \psi_{0,\phi\phi} + 2iS_{1\phi}\psi_{0,\phi}\Big]- S_{2,z}\psi_0 \\& \hspace{0.4 in}+{1\over z}
\left(\frac{3\alpha_0}{\phi^2}-\frac{x^2\alpha_0\beta_0}{z^2\phi^2}+\frac{2x^2\beta_1\phi}{z^2}\right)S_{2,\phi} = 0.\label{semi3}\end{split}\ee
To find $S_0,\; S_1$ and $S_2$ etc., the above set of equations \eqref{semi1} - \eqref{semi3} should be solved sequentially. Now identifying $S_{0,x}$ as $p_x$, $S_{0,\phi}$ as $p_\phi$, $S_{0,z}$ as $p_z$; the classical Hamiltonian constraint equation ${\mathcal{H}} = 0$, presented in equation (\ref{Hp1c}) may easily be retrieved from equation (\ref{semi1}). Thus (\ref{semi1}) is recognized as the Hamilton-Jacobi equation, and hence the Hamilton-Jacobi function, $S_0(x,z,\phi)$ is expressed as,

\be\label{S0} S_0 = \int p_x dx + \int p_z dz + \int p_\phi d\phi, \ee
where we absorb the constant of integration in $\psi_0$. The integrals in the above expression can be evaluated using the classical solution for $k = 0$ presented in equation (\ref{param}), and also using the expressions for $p_x,~p_z,~p_{\phi}$ given in (\ref{p1}) and (\ref{pz}). Further, it is required to use the relation, $x = \dot z ~(N = 1)$, where, $z = a^2$, and to choose $n = -5$, since probability interpretation holds only for such value of $n$. As a result, using solution (\ref{param}), we obtain the following set of relations in which the momenta $p_x$, $p_z$ $p_\phi$ are now expressed in term of $x$, $z$ and $\phi$ respectively:

\begin{subequations}\begin{align}
&\label{allp} ~~~~~~~~~~~~~~~~~~~~~~~~~~~~~~~~~~~~\alpha'=-\left(\frac{\alpha_0}{\phi^2}\right);\\
&~~~~~~~~~~~~~~~~~~~~~~~~~~~~~~~~~~~\beta'=-\frac{\phi}{24\lambda^2}+\frac{\alpha_0}{\phi^2}\left[\frac{1}{2\lambda^2}-\frac{\Lambda}{12\lambda^4}\right];\\
&~~~~~~~~~~~~~~~~~~~~~~~~~~~~~~~~~~~ x = 2{\lambda} z ;\\
&~~~~~~~~~~~~~~~~~~~~~~~~~~~~~~~~~~~p_x =576\sqrt2\gamma_0{\lambda}^{\frac{11}{2}} x^{\frac{1}{2}};\\
&~~~~~~~~~~~~~~~~~~~~~~~~~~~~~~~~~~~p_z =-\frac{10368\gamma_0\lambda^7\sqrt z}{7}-\frac{3\lambda\alpha_0 z}{a_0\phi_0}-\frac{a_0^2\phi_0^2\lambda}{2\sqrt z}-\frac{\alpha_0z\Lambda}{\lambda a_0\phi_0} ;\\
&~~~~~~~~~~~~~~~~~~~~~~~~~~~~~~~~~~~p_\phi =\frac{2\alpha_0{\lambda} a_0^3\phi_0^3}{\phi^5}-\frac{2\lambda a_0^3\phi_0^3}{3\phi^2} + \frac{2\alpha_0\Lambda a_0^3\phi_0^3} {3\lambda\phi^5}.
                                \end{align}\end{subequations}
The integrals appearing in (\ref{S0}) can now be evaluated in a straightforward manner as,
\begin{subequations}\begin{align}
&\label{px2pphi}~~~~~~~~~~~~~~~~~~~~~~~~~~~~~~~~~~~~~~
\int p_x dx =384\sqrt2\gamma_0{\lambda}^{\frac{11}{2}} x^{\frac{3}{2}}, \\
&~~~~~~~~~~~~~~~~~~~~~~~~~~~~~~~~~~~~~~~~\int p_z dz=-\frac{6912\gamma_0\lambda^7 z^{3\over 2}}{7}-\frac{3\lambda\alpha_0 z^2}{2a_0\phi_0}-{a_0^2\phi_0^2\lambda\sqrt z}-\frac{\Lambda\alpha_0 z^2}{2\lambda a_0\phi_0}, \\
&~~~~~~~~~~~~~~~~~~~~~~~~~~~~~~~~~~~~~~~~\int p_\phi d\phi = -\frac{\alpha_0{\lambda} a_0^3\phi_0^3}{2\phi^4}+\frac{2\lambda a_0^3\phi_0^3}{3\phi}- \frac{\alpha_0\Lambda a_0^3\phi_0^3}{6\lambda\phi^4},
\end{align}\end{subequations}
and explicit form of the function $S_0$ in terms of $z$ is found as,

\be\label{S0f}~~~~~~~~~~~~~~~~~~~~~~~~~~~~~~~~~~~~ S_0 =\frac{3840\gamma_0 \lambda^7z^{3\over 2}}{7}-\frac{2\lambda\alpha_0 z^2}{a_0\phi_0}-\frac{2\alpha_0\Lambda z^2}{3\lambda a_0\phi_0}-\frac{\lambda a_0^2\phi_0^2 \sqrt z}{3} .\ee
One can now quite easily check that the expression for the function $S_0$ so obtained in equation (\ref{S0f}) satisfies equation (\ref{S0}) identically. In fact it must, because equation (\ref{S0}) is identified with Hamiltonian constraint equation (\ref{Hp1c}) for $k = 0$. Thus, $S_0$ is identified unambiguously as the Hamilton-Jacobi function. Moreover, it is possible to compute on-shell action to the zeroth order (\ref{A2}) in the following manner. Using equation (\ref{aphi}) and  the classical solution (\ref{param}), it is possible to express all the variables in terms of $t$ and substitute in the action (\ref{A2}) to obtain,

\be\label{Acl}A=A_{cl}=\int\left[\frac{11520\gamma_0{a_0}^{3}{\lambda}^{8}e^{3\lambda t}}{7}-\frac{8\alpha_0\lambda^2 a_0^3}{\phi_0}e^{{4\lambda}t}-\frac{4\alpha_0a_0^3\Lambda e^{4{\lambda}t}}{3\phi_0}-\frac{\lambda ^2a_0^3\phi_0^2 e^{{\lambda}t}}{3} \right]dt. \ee
Integrating we have,

\be\label{4.87} A=A_{cl}=\frac{3840\gamma_0{a_0}^{3}{\lambda}^{7}e^{3\lambda t}}{7}-\frac{2\alpha_0 a_0^3\lambda e^{{4\lambda}t}}{\phi_0}-\frac{\alpha_0a_0^3\Lambda e^{4{\lambda}t}}{3\lambda\phi_0}-\frac{\lambda a_0^3\phi_0^2}{3} e^{{\lambda}t}.\ee
Since $\sqrt z = a = a_0e^{\lambda t}$, therefore the classical on-shell action \eqref{4.87} is identical to the Hamilton-Jacobi function (\ref{S0f}), and everything so far, is well behaved and consistent. The wave function, upto zeroth order approximation now reads as,

\be\label{psif} \Psi = \psi_0 e^{\frac{i}{\hbar}\left[ \frac{3840\gamma_0{a_0}^{3}{\lambda}^{7}e^{3\lambda t}}{7}-\frac{2\alpha_0 a_0^3\lambda e^{{4\lambda}t}}{\phi_0}-\frac{\alpha_0a_0^3\Lambda e^{4{\lambda}t}}{3\lambda\phi_0}-\frac{\lambda a_0^3\phi_0^2}{3} e^{{\lambda}t}\right]}.\ee
Solving the first order equation (\ref{semi2}) exactly is onerous. Nevertheless, since we have expressed all the variables in terms of $z$, so one can neglect some derivative terms associated with slowly varying amplitude $\psi_0$, and perform a little algebra, to reveal the fact that one can in principle, express (\ref{semi2}) in the form $S_1 = iF_1(z)$ on the solutions (\ref{param}). Therefore  the wavefunction, up to the first order approximation may be expressed as:
\begin{center}
\be\label{psi} \Psi =\psi_{01}e^{\frac{i}{\hbar}}\left[\frac{3840\gamma_0{a_0}^{3}{\lambda}^{7}e^{3\lambda t}}{7}-\frac{2\alpha_0 a_0^3\lambda e^{{4\lambda}t}}{\phi_0}-\frac{\alpha_0a_0^3\Lambda e^{4{\lambda}t}}{3\lambda\phi_0}-\frac{\lambda a_0^3\phi_0^2}{3} e^{{\lambda}t}\right] .\ee
\end{center}
where,

\be \Psi_{01}=\Psi_0e^{-F(z)}.\ee
Clearly, up to first order approximation of the semiclassical wavefunction \eqref{psi}, only the pre-factor is modified, while the  oscillatory behaviour of the wave function remains intact. Indeed it is possible to proceed further in order to evaluate the semiclassical wavefunction under higher order approximations. Nonetheless, the form (exponent) of the semiclassical wavefunction remains unaltered. Such, exhibition of oscillatory behaviour infers that the wave-function is firmly peaked around the classical inflationary solutions \eqref{param}. Hence Hartle prescription \cite{Hartle} regarding the emergence of classical trajectory is ensured. Such, a smooth emergence from quantum domain to the classical universe we live in, validates the present model under consideration, in the context of very early evolution of the universe.

\section{Appendix B: Phase space structure, Modified Horowitz' formalism:}
As mentioned earlier in subsection (3), unlike GTR, time derivative of the lapse function $N$ appears in the action \eqref{A2}. The appearance of $\dot N^2$ term in the action is definitely uncanny, since it behaves like a dynamical variable. On the contrary, the Hessian determinant of the vanishes, implying that the action is degenerate. The lapse function is already known to be simply a gauge, which is responsible for such degeneracy. Thus establishing diffeomorphic invariance is non-trivial for higher-order theory. To remove degeneracy, we follow Dirac's algorithm of constraint analysis in Appendix A, which provides correct Hamiltonian if the action is cast in terms of the induced three metric and divergent terms are taken into account a-priori. There exists yet an alternative technique to construct the Hamiltonian bypassing Dirac's programme. This is known as Modified Horowitz' Formalism (MHF). In this formalism, the action again is expressed in terms of the induced three metric, and the divergent terms are removed. Thereafter, a canonical auxiliary variables is introduced. At the end one requires to switch over from the auxiliary variable to the basic variable $K_{ij}$ through a canonical transformation. For the sake of completeness, in this Appendix we show that the MHF leads to the identical Hamiltonian\eqref{Hp1c}. It is important to mention that, there exists several other formalisms, which produce canonically equivalent Hamiltonian. However, such canonical transformation cannot be extended in the quantum domain, due to non-linearity. As a result, all other techniques produce different quantum dynamics.\\

In MHF, the first step is to express the action in terms of the basic variables $h_{ij}$. Next, the total derivative terms are removed under integration by parts. These terms are canceled with the supplementary boundary terms. We therefore start from the divergent free action \eqref{A2}, to find the canonical auxiliary variables, taking the derivative of the action \eqref{A2} with respect to the highest derivative present in it. Therefore, taking into account the supplementary boundary terms \bigg[$\alpha(\phi)\Sigma_{R}=-\frac{3\alpha \dot z\sqrt z}{N},~~\beta(\phi)\Sigma_{\mathcal{G}} =-\beta(\phi){\dot z \over N {\sqrt z}}\left({\dot z^2 \over N^2 z }+12 K\right)$ and~ $\gamma(\phi)\Sigma_{\mathcal{G}_1^2}= \frac{9\gamma\dot z^7}{7N^7z^{11\over 2}}+\frac{72k\gamma \dot z^5}{5N^5z^{9\over 2}}+\frac{48k^2\gamma\dot z^3}{N^3z^{7\over 2}}$\bigg], which cancel the divergent terms, auxiliary variable

\be Q=\frac{\partial A}{\partial \ddot{z}}=144\gamma\bigg[\frac{2\ddot{z}}{16N^3z^\frac{9}{2}}\Big{(}\frac{\dot{z}^2}{N^2}+4kz\Big{)}^2-\Big({\dot N\dot z^5\over 8 N^8 z^{9\over 2}}+{k\dot N\dot z^3\over N^6 z^{7\over 2}}+{2k^2\dot N\dot z\over N^4 z^{5\over 2}}\Big)\bigg]\ee
is introduced straight into the action \eqref{A2} as

\be\begin{split} \label{a3} A = \int &\Bigg[{6\alpha N}\Big(k\sqrt z-\frac{\dot{z}^2}{4N^2\sqrt z}-\frac{\Lambda z^{3\over 2}}{3}\Big)-\frac{3\alpha' \dot\phi \dot z\sqrt z}{N}-\frac{\beta'\dot z\dot\phi}{N\sqrt z}\bigg{(}\frac{\dot z^2}{N^2 z}+12k \bigg{)}+ N z^{3\over 2}\Big({\dot\phi^2\over 2 N^2}-V\Big)\\&+\Big(Q\ddot{z}-\frac{16Q^2N^3z^\frac{9}{2}}{576\gamma\Big(\frac{\dot{z}^2}{N^2}+4kz\Big)^2}-\frac{Q\dot{N}\dot{z}}{N}\Big) + 144\gamma\Big\{-\Big( {15\dot z^8 \over 448 N^7 z^{13\over 2}}+{13 k\dot z^6\over 40 N^5z^{11\over 2}}+{11 k^2\dot z^4\over 12 N^3z^{9\over 2}}\Big)\\&+{\gamma'\dot\phi\over \gamma}\Big({\dot z^7\over 112 N^7 z^{11\over 2}}+{k\dot z^5\over 10 N^5 z^{9\over 2}}+{k^2\dot z^3\over 3 N^3 z^{7\over 2}}\Big)\Big\}\Bigg]dt+\gamma(\phi)\Sigma_{\mathcal{G}_2^2}.\end{split}\ee
Integrating the action by parts, the total derivative term $Q\dot z$ is canceled with the supplementary boundary term [$\gamma(\phi)\Sigma_{\mathcal{G}_2^2}=-Q\dot z$], and the resulting action becomes canonical, which reads as

\be\begin{split} \label{A4} A = \int &\Bigg[{6\alpha N}\Big(k\sqrt z-\frac{\dot{z}^2}{4N^2\sqrt z}-\frac{\Lambda z^{3\over 2}}{3}\Big)-\frac{3\alpha' \dot\phi \dot z\sqrt z}{N}-\frac{\beta'\dot z\dot\phi}{N\sqrt z}\bigg{(}\frac{\dot z^2}{N^2 z}+12k \bigg{)}\\&+ N z^{3\over 2}\Big({\dot\phi^2\over 2 N^2}-V\Big)-\Big(\dot{Q}\dot{z}+\frac{16Q^2N^3z^\frac{9}{2}}{576\gamma\Big(\frac{\dot{z}^2}{N^2}+4kz\Big)^2}+\frac{Q\dot{N}\dot{z}}{N}\Big)\\& + 144\gamma\Big\{-\Big( {15\dot z^8 \over 448 N^7 z^{13\over 2}}+{13 k\dot z^6\over 40 N^5z^{11\over 2}}+{11 k^2\dot z^4\over 12 N^3z^{9\over 2}}\Big)+{\gamma'\dot\phi\over \gamma}\Big({\dot z^7\over 112 N^7 z^{11\over 2}}+{k\dot z^5\over 10 N^5 z^{9\over 2}}+{k^2\dot z^3\over 3 N^3 z^{7\over 2}}\Big)\Big\}\Bigg]dt.\end{split}\ee
Therefore, the canonical momenta are

\be\begin{split}\label{momenta} &p_Q=-\dot{z},\\& p_z=-\frac{3\alpha\dot{z}}{N\sqrt{z}}-\frac{3\alpha^{\prime}\sqrt{z}\dot{\phi}}{N}-\frac{\beta'\dot\phi}{N\sqrt z}\bigg{(}\frac{\dot z^2}{N^2 z}+12k \bigg{)}-\dot{Q}+\frac{64Q^2Nz^\frac{9}{2}\dot{z}}{576\gamma\Big[\frac{\dot{z}^2}
{N^2}+4kz\Big]^3}-\frac{Q\dot{N}}{N}\\&-144\gamma\bigg[{120\dot z^7 \over 448 N^7 z^{13\over 2}}+{78 k\dot z^5\over 40 N^5z^{11\over 2}}+{44 k^2\dot z^3\over 12 N^3z^{9\over 2}}\bigg]+144{\gamma'\dot\phi}\bigg[{7\dot z^6\over 112 N^7 z^{11\over 2}}+{k\dot z^4\over 2 N^5 z^{9\over 2}}+{k^2\dot z^2\over  N^3 z^{7\over 2}}\bigg],\\& p_\phi=-\frac{3\alpha'\sqrt{z}\dot{z}}{N}-\frac{\beta'\dot z}{N\sqrt z}\bigg{(}\frac{\dot z^2}{N^2 z}+12k \bigg{)}+\frac{z^\frac{3}{2}\dot{\phi}}{N}+144\gamma'\bigg[{\dot z^7\over 112 N^7 z^{11\over 2}}+{k\dot z^5\over 10 N^5 z^{9\over 2}}+{k^2\dot z^3\over 3 N^3 z^{7\over 2}}\bigg],\\& p_N=-\frac{Q\dot{z}}{N}. \end{split}\ee
The action \eqref{A4} is not free from the time derivative of the Lapse function $N$ and also all the momenta are not invertible. This clearly imply that the action is degenerate. However as mentioned, instead of Dirac’s constraint analysis, here we proceed to construct the Hamiltonian in the following unique manner. Let us start from the following expression

\be\begin{split}p_Qp_z=&\frac{3\alpha\dot{z}^2}{N\sqrt{z}}+\frac{3\alpha^{\prime}\sqrt{z}\dot{z}\dot{\phi}}{N}+\frac{\beta'\dot\phi\dot z}{N\sqrt z}\bigg{(}\frac{\dot z^2}{N^2 z}+12k \bigg{)}+\dot{Q}\dot{z}-\frac{64Q^2Nz^\frac{9}{2}\dot{z}^2}{576\gamma\Big[\frac{\dot{z}^2}
{N^2}+4kz\Big]^3}+\frac{Q\dot{N}\dot{z}}{N}\\&+144\gamma\bigg[{120\dot z^8 \over 448 N^7 z^{13\over 2}}+{78 k\dot z^6\over 40 N^5z^{11\over 2}}+{44 k^2\dot z^4\over 12 N^3z^{9\over 2}}\bigg]-144{\gamma'\dot\phi}\bigg[{7\dot z^7\over 112 N^7 z^{11\over 2}}+{k\dot z^5\over 2 N^5 z^{9\over 2}}+{k^2\dot z^3\over N^3 z^{7\over 2}}\bigg],\end{split}\ee
obtained in view of the first two relations of equation \eqref{momenta}, and construct the Hamiltonian in terms of the phase space variables as,

\be\label{H1h}\begin{split}  H_{\mathcal{MH}}=\bigg{[}&-p_Q p_z+{N^3Q^2\over 576\gamma \left({p_Q^4\over 16N^4 z^{9\over 2}}+{k p_Q^2\over 2 N^2 z^{7\over 2}}+{k^2\over z^{5\over 2}}\right)}+{N p_{\phi}^2\over 2z^{3\over 2}}-{3\alpha' p_Q p_{\phi}\over z}-{\beta' p_Q p_{\phi}\over z^2}\left({\frac{p_Q^2}{N^2 z}}+12k\right)\\&+{144\gamma' N p_{\phi}\over z^{3\over 2}}\left({p_Q^7\over 112 N^7 z^{11\over 2}} + {k p_Q^5\over 10 N^5 z^{9\over 2}}+ {k^2 p_Q^3\over 3 N^3 z^{7\over 2}}\right)+144\gamma \left({15p_Q^8\over 448 N^7 z^{13\over 2}}+{13k p_Q^6\over 40N^5z^{11\over 2}}+{11k^2p_Q^4\over 12 N^3 z^{9\over 2}}\right)\\&+\frac{3\alpha'\beta'p_Q^2}{Nz^{3\over 2}}\left({\frac{p_Q^2}{N^2 z}}+12k\right)-{432\alpha'\gamma' \over z}\left({p_Q^8\over 112 N^7 z^{11\over 2}} + {k p_Q^6\over 10 N^5 z^{9\over 2}}+ {k^2 p_Q^4\over 3 N^3 z^{7\over 2}}\right)\\&-{288\beta'\gamma' p_Q\over z}\bigg{(}\frac{p_Q^2}{N z}+12k \bigg{)}\left({p_Q^7\over 112 N^7z^{11\over 2}} + {k p_Q^5\over 10 N^5 z^{9\over 2}}+ {k^2 p_Q^3\over 3N^3 z^{7\over 2}}\right)+{9\alpha'^2 p_Q^2\over 2N\sqrt z}\\&+{\beta'^2 p_Q^2\over 2Nz^{3\over 2}}\bigg{(}\frac{p_Q^2}{N^2z}+12k \bigg{)}^2+{10368\gamma'^2\over z^{3\over 2}}\left({p_Q^7\over 112 N^7 z^{11\over 2}} + {k p_Q^5\over 10 N^5 z^{9\over 2}}+ {k^2 p_Q^3\over 3 N^3 z^{7\over 2}}\right)^2\\&+ {6\alpha N}\left({p_Q^2\over 4 N^2 \sqrt z} - k\sqrt z + {\Lambda z^{3\over 2}\over 3}\right)+NVz^{3\over 2}\Bigg{]}.\end{split}\ee
Now to establish diffeomorphic invariance we need to switch over to the basic variables $(x, z, \phi)$ and their canonically conjugate momenta $(p_x, p_z, p_{\phi})$. For this purpose, we make the canonical transformations, replacing $Q=\frac{p_x}{N}$ and $p_Q=-Nx$. The Hamiltonian in terms of the basic variables is then found as,

\be\label{H2h}\begin{split} H_{\mathcal{MH}}=N\Bigg{[}&x p_z+{p_x^2\over 576\gamma \left({x^4\over 16 z^{9\over 2}}+{k x^2\over 2 z^{7\over 2}}+{k^2\over z^{5\over 2}}\right)}+{p_{\phi}^2\over 2z^{3\over 2}}+{3\alpha' x p_{\phi}\over z}+\frac{\beta'x p_{\phi}}{ z^2}\bigg{(}\frac{x^2}{z}+12k \bigg{)}\\&-{144\gamma'p_{\phi}\over z^{3\over 2}}\left({x^7\over 112 z^{11\over 2}} + {k x^5\over 10 z^{9\over 2}}+ {k^2 x^3\over 3 z^{7\over 2}}\right)+144\gamma \left({15x^8\over 448 z^{13\over 2}}+{13k x^6\over 40z^{11\over 2}}+{11k^2 x^4\over 12 z^{9\over 2}}\right)\\&+{3\alpha'\beta' x^2\over z^{3\over 2}}\bigg{(}\frac{x^2}{z}+12k \bigg{)}-{432\alpha'\gamma' \over z}\left({x^8\over 112 z^{11\over 2}} + {k x^6\over 10 z^{9\over 2}}+ {k^2 x^4\over 3 z^{7\over 2}}\right)\\&-{288\beta'\gamma' x\over z}\bigg{(}\frac{x^2}{z}+12k \bigg{)}\left({x^7\over 112 z^{11\over 2}} + {k x^5\over 10 z^{9\over 2}}+ {k^2 x^3\over 3 z^{7\over 2}}\right)+{9\alpha'^2 x^2\over 2\sqrt z}\\&+{\beta'^2 x^2\over 2z^{3\over 2}}\bigg{(}\frac{x^2}{z}+12k \bigg{)}^2+{10368\gamma'^2\over z^{3\over 2}}\left({x^7\over 112 z^{11\over 2}} + {k x^5\over 10 z^{9\over 2}}+ {k^2 x^3\over 3 z^{7\over 2}}\right)^2\\&+ {6\alpha}\left({x^2\over 4\sqrt z} - k\sqrt z + {\Lambda z^{3\over 2}\over 3}\right)+Vz^{3\over 2}\Bigg{]}=N\mathcal{H}.\end{split}\ee
The Hamiltonian \eqref{H2h} so obtained is clearly the same as the one obtained following Dirac algorithm \eqref{Hp1c}. This establishes that the computation carried out with Dirac's formalism in the Appendix A is rigorous.

\section{Appendix C: Hermiticity and the unitarity of $\widehat H_e$:}

To prove that the effective Hamiltonian operator (\ref{qh2})is hermition, let us split $\widehat H_e$ as,

\begin{center}
\be\label{He}\begin{split} &\widehat H_e= -\frac{\hbar^2}{198\gamma x^5}\bigg{(}\frac{\partial^2}{\partial x^2} +\frac{n}{x}\frac{\partial}{\partial x}\bigg{)}-\frac{\hbar^2}{11x\sigma^{\frac{12}{11}}}\frac{\partial^2}{\partial \phi^2}+\frac{6i\hbar\alpha_0} {11\sigma}\bigg{(}\frac{1}{\phi^2}\frac{\partial}{\partial \phi}-\frac{1}{\phi^3}\bigg{)}\\&-\frac{2i\hbar x^2\alpha_0 \beta_0}{11\sigma^{15\over 11}}\bigg{(}\frac{1}{\phi^2}\frac{\partial}{\partial \phi}-\frac{1}{\phi^3}\bigg{)}+\frac{2i\hbar x^2\beta_1}{11\sigma^{15\over 11}}\bigg{(}2\phi\frac{\partial}{\partial \phi}+1\bigg{)}+V_{e}=\widehat H_1+\widehat H_2+\widehat H_3+\widehat V_{e},\end{split}\ee
\end{center}
where,
\begin{center}
\begin{eqnarray}
\widehat H_1& =& -\frac{\hbar^2}{198\gamma x^5}\bigg{(}\frac{\partial^2}{\partial x^2} +\frac{n}{x}\frac{\partial}{\partial x}\bigg{)} \\
\widehat H_2& =& -\frac{\hbar^2}{11x\sigma^{\frac{12}{11}}}\frac{\partial^2}{\partial \phi^2} \\
\widehat H_3&=&\frac{6i\hbar\alpha_0} {11\sigma}\bigg{(}\frac{1}{\phi^2}\frac{\partial}{\partial \phi}-\frac{1}{\phi^3}\bigg{)}-\frac{2i\hbar x^2 \alpha_0\beta_0}{11\sigma^{15\over 11}}\bigg{(}\frac{1}{\phi^2}\frac{\partial}{\partial \phi}-\frac{1}{\phi^3}\bigg{)}+\frac{2i\hbar x^2\beta_1}{11\sigma^{15\over 11}}\bigg{(}2\phi\frac{\partial}{\partial \phi}+1\bigg{)}\\
\widehat V_{e}& = &V_{e}.
\end{eqnarray}
\end{center}
Now, let us consider the first term,

\begin{center}
\be \label{46} \int \big{(}\widehat H_1\Psi\big{)}^*\Psi dx= -\frac{\hbar^2}{198}\int\frac{1}{\gamma x^5}\bigg{(}\frac{\partial^2\Psi^*}{\partial x^2} +\frac{n}{x}\frac{\partial\Psi^*}{\partial x}\bigg{)}\Psi dx.\ee
\end{center}
Under integration by parts twice and dropping the first term due to fall-of condition, we obtain,
\be\label{H1}\begin{split} \int \big{(}\widehat H_1\Psi\big{)}^*\Psi dx=-\frac{\hbar^2}{198} \int \Psi^*\Bigg{[}\frac{1}{{x^5}\gamma }\frac{\partial^2\Psi}{\partial x^2} - \frac{{(n+10)x^4}\gamma} {\left[ {x^5}\gamma \right]^2} \frac{\partial \Psi}{\partial x}\Bigg] + \frac{\hbar^2}{198} \int\Psi^*\Psi\frac{\partial }{\partial x}\Bigg(\frac{{x^4}\gamma\left(n+5\right)}{\left[{x^5}\gamma\right]^2 } \Bigg) dx. \end{split}\ee
To proceed further, one has to get rid of the last term appearing in equation (\ref{H1}). The only choice for which last term in the above expression vanishes is, $n=-5$. Therefore we have,

\begin{center}
\be\label{H1.1} \int \big{(}\widehat H_1\Psi\big{)}^*\Psi dx=-\frac{\hbar^2}{198\gamma}\int \Psi^*\bigg{[}\frac{1}{x^5}\frac{\partial^2\Psi}{\partial x^2} -\frac{5}{x^6}\frac{\partial\Psi}{\partial x}\bigg{]}dx=\int \Psi^*\widehat H_1\Psi dx. \ee
\end{center}
Thus $\widehat H_1$ is hermitian, for a particular choice of operator ordering parameter $n=-5$. Further, since $\widehat H_2$ is typically hermitian, therefore, let us take up $\widehat H_3$, next.

\be \begin{split}\int (\widehat H_3 \Psi)^*\Psi d\phi=& -\frac{6i\hbar\alpha_0 } {11\sigma}\int\bigg{(}\frac{1}{\phi^2}\frac{\partial\Psi^*}{\partial \phi}\Psi-\frac{1}{\phi^3}\Psi^*\Psi \bigg{)} d\phi +\frac{2i\hbar x^2\alpha_0\beta_0}{11\sigma^{15\over 11}}\int \bigg{(}\frac{1}{\phi^2}\frac{\partial\Psi^*}{\partial \phi}\Psi-\frac{1}{\phi^3}\Psi^*\Psi \bigg{)}d\phi\\&-\frac{2i\hbar x^2\beta_1}{11\sigma^{15\over 11}}\int\bigg{(}{2\phi}\frac{\partial\Psi^*}{\partial \phi}\Psi+\Psi^*\Psi \bigg{)}d\phi.\end{split}\ee
Integrating by parts and dropping the terms due to fall-of condition, we obtain,

\be\begin{split} \int (\widehat H_3 \Psi)^*\Psi d\phi= &\frac{6i\hbar\alpha_0 }{11\sigma}\int\Psi^* \bigg{(}\frac{1}{\phi^2}\frac{\partial\Psi}{\partial \phi}-\frac{1}{\phi^3}\Psi \bigg{)}  d\phi -\frac{2i\hbar x^2\alpha_0\beta_0}{11\sigma^{15\over 11}}\int \Psi^*\bigg{(}\frac{1}{\phi^2}\frac{\partial\Psi}{\partial \phi}-\frac{1}{\phi^3}\Psi\bigg{)} d\phi\\&+\frac{2i\hbar x^2\beta_1}{11\sigma^{15\over 11}}\int \Psi^*\bigg{(}{2\phi}\frac{\partial\Psi}{\partial \phi}+\Psi\bigg{)}=\int \Psi^*\widehat H_3 \Psi d\phi,\end {split}\ee
and hence, $\widehat H_3$ is also a hermitian operator, and as a result the effective Hamiltonian operator $\widehat H_e$ turns out to be a hermitian operator too.\\

Let us now express $\widehat H_0 = \widehat H_1 + \widehat H_2$, and $\widehat H_I = \widehat H_3 + \widehat V_e$, where $\widehat H_I$ is the interacting term. Now $\widehat H_0$ being hermitian, it is solvable, and the eigen-decomposition are known. Thus, one can find the propagator as $U_0 = e^{-{i\over \hbar}\widehat H_0\tau}$. Now, as $\widehat H_I$ is also hermitian, one may be tempted to write $U_I = e^{-{i\over \hbar}\int_0^t \widehat H_I(\tau)d\tau}$. But the problem is $\widehat H_I$ does not commute at two different epochs, i.e. $[\widehat H_I(\tau_1), \widehat H_I(\tau_2)] \ne 0$. We therefore, cannot find explicit solution in terms of an integral. Nonetheless, the propagator in Dyson interacting picture takes the form,

\be U_I = \mathcal{T} e^{-{i\over \hbar}\int _{\tau_0} ^\tau H_I d\tau},\ee
where, $\mathcal{T}$ is the time ordering operator. Hence we find the unitary operator of the time-dependent Hamiltonian under consideration. The unitarity of the effective Hamiltonian operator is therefore tentatively established.

\end{document}